\documentclass[english]{tlp}
\usepackage[latin1]{inputenc}
\usepackage{babel}
\usepackage{graphics}
\usepackage{graphicx}
\usepackage{aopmath}
\usepackage{amsfonts}
\usepackage{amssymb}
\usepackage{color}
\usepackage{multirow}

\newtheorem{example}{Example}         % [section]
\newtheorem{definition}{Definition}   % [section]
\usepackage{algorithmicx}
\usepackage{algpseudocode}
\usepackage[ruled,vlined]{algorithm2e}
\usepackage{enumitem} 
\usepackage{listings}
\usepackage{url}
\usepackage{tikz}   
\renewcommand{\setminus} 
{\hbox{\hspace{1pt}\tikz{\draw[line width=.7pt,line cap=round] (3pt,0) -- (0,6pt);}}\hspace{.5pt}} 
\SetKwInput{KwInput}{Input}
\SetKwInput{KwOutput}{Output}

\algnewcommand{\LeftComment}[1]{\Statex $\hspace*{-2.5mm} \bullet$ #1}

\newcommand{\clspace}{\hspace{5mm}}

\usepackage{color}

\newcommand{\exm}

\title{Predicate Pairing for Program Verification}

\author[E. De~Angelis, F. Fioravanti, A. Pettorossi, M. Proietti]
{EMANUELE DE ANGELIS, FABIO FIORAVANTI\\
DEC, University `G. d'Annunzio', Pescara, Italy\\
\email{\{emanuele.deangelis,fabio.fioravanti\}@unich.it}
\and
ALBERTO PETTOROSSI\\
DICII, Universit\`a di Roma Tor Vergata, Roma, Italy\\
\email{pettorossi@info.uniroma2.it}
\and
MAURIZIO PROIETTI\\
CNR-IASI, Roma, Italy\\
\email{maurizio.proietti@iasi.cnr.it} 
}

\submitted{}
\revised {}
\accepted{}

\begin{document}

% \vspace{-2mm}
%\pagestyle{plain}
%\pagestyle{empty} %no page numbers

\maketitle

\vspace{-4mm}
\noindent
{\bf{Note:}} This article has been published in 
{\it{Theory and Practice of Logic Programming}}, 18(2), 126--166,
\copyright Cambridge University Press.

\vspace{2mm}

\begin{abstract}
It is well-known that the verification of
partial correctness properties of imperative programs can be
reduced to the satisfiability
problem for constrained Horn clauses (CHCs). 
However, state-of-the-art solvers for constrained
Horn clauses (or CHC solvers) based on {\it predicate abstraction}
are sometimes unable to verify satisfiability because they look for models 
that are definable in a given class $\mathcal A$ of constraints, called $\mathcal A$-definable models. 
We introduce a transformation technique, called {\it Predicate Pairing}, 
which is able, in many interesting cases, to transform a set of clauses into an equisatisfiable
set whose satisfiability can be 
proved by finding an $\mathcal A$-definable model, and hence
can be effectively verified by a state-of-the-art CHC solver.

In particular, we prove that, under very general conditions on $\mathcal A$,
the unfold/fold transformation 
rules preserve the existence of an $\mathcal A$-definable 
model, that is,
if the original clauses have an $\mathcal A$-definable model, 
then the transformed clauses have an $\mathcal A$-definable model.
The converse does not hold in general, and we 
provide suitable conditions under which 
the transformed clauses have an $\mathcal A$-definable model 
{\it if and only if}
the original ones have an $\mathcal A$-definable model. 
Then, we present a strategy, called Predicate Pairing, which guides the
application of the transformation rules with the objective of deriving
a set of clauses whose satisfiability problem can be solved by looking
for $\mathcal A$-definable models.
The Predicate Pairing strategy introduces
a new predicate defined by the conjunction
of two predicates occurring in the original set of clauses,
together with a conjunction of constraints. 
We will show through some examples that an $\mathcal A$-definable model
may exist for the new predicate even if it does not exist for its 
defining atomic conjuncts.
We will also present some case studies showing that
Predicate Pairing plays a crucial role in the verification of 
{\it relational properties of programs}, that is, 
properties relating two programs (such as program
equivalence) or two executions of the same program (such as 
non-interference).
Finally, we perform an experimental evaluation of the proposed techniques 
to assess the effectiveness of Predicate Pairing 
in increasing the %as a way of improving the 
power of CHC solving.

\noindent
%\textit{Under consideration for publication in Theory and Practice of Logic Programming} (TPLP).
\end{abstract}

\begin{keywords}
Program Verification,
Constrained Horn Clauses,
Constraint Logic Programming,
Program Transformation,
Relational Properties of Programs\nopagebreak
\end{keywords}\nopagebreak

%\vspace{-2mm}
\section{Introduction}
\label{sec:intro}
{\it Constrained Horn clauses} (CHCs, for short) have been advocated 
by many researchers as a suitable logical formalism
for the specification and the automated verification of properties
of imperative
programs~\cite{Al&07,Bj&15,De&14c,Ja&09,Ka&16,Me&07,Pe&98,PoR07,Ru&13}.
In particular, the problem of showing {\it partial correctness} properties defined by Hoare triples~\cite{Hoa69} has a natural translation into the problem of proving
the {\it satisfiability} of a suitable set of constrained Horn clauses.

Consider, for instance, the C-like program \texttt{sum\_upto} in 
Figure~\ref{fig:sumupto}, which computes the sum of the first {\tt m}
non-negative integer numbers:

\renewcommand{\baselinestretch}{.9}
\begin{figure}[h]
\centering
\vspace{-1mm}
%\vline ~ 
\begin{minipage}[t]{0.4\textwidth}
\begin{verbatim}
int m, sum;                                       
int f(int x) {                                      
   int r = 0;
   while (x > 0) { 
      r = r + x;  x--; }
   return r; 
   }
\end{verbatim}
\end{minipage}

\vspace{2mm}
%\vline ~ 
\begin{minipage}[t]{0.4\textwidth}
\begin{verbatim}
void sum_upto() {
   sum = f(m); 
   }
\end{verbatim}
\end{minipage} 
% \vline
%\vspace*{-1mm}
\caption{\label{fig:sumupto} Program \texttt{sum\_upto} computing 
${\mathtt{sum}} =\sum_{{x=1}}^{\mathtt m}x$.}
\vspace{-3mm}
\end{figure}

\renewcommand{\baselinestretch}{1}
Suppose we want to prove the following Hoare triple:
\mbox{$\{ {\tt m}\!\geq\!0\} \ \texttt{sum\_upto} \  \{ {\tt sum}\!\geq\! {\tt m}\}$.}
This triple is valid if the following set of clauses, called 
{\it verification conditions}, is satisfiable:

\smallskip
\noindent
1. \ $\textit{false} \leftarrow  M\!>\!\textit{Sum},\ M\!\geq\! 0,\ R\!=\!0,\ 
\textit{su\/}(M\!,R,\textit{Sum})$	

\noindent
2. \ $\textit{su\/}(X\!,R,\textit{Sum}) \leftarrow X\!\leq\!0,\ \textit{Sum}=\!R$

\noindent
3. \ $\textit{su\/}(X\!,R,\textit{Sum}) \leftarrow X\!>\!0,\ 
R1\!=\!R\!+\!X,\ X1\!=\!X\!-\!1,\ \textit{su\/}(X1,R1,\textit{Sum})$

\smallskip

%\comment{REVIEWER: a full set of references should be given for previous work on the 
%generation of verification conditions, rather than an indirect reference through a previous 
%paper by the authors.}
These clauses can be obtained in an automatic way from an interpreter of
the \mbox{C-like} imperative language we consider and the given Hoare triple by using
a technique described in the literature~\cite{Al&07,De&14c,De&17b,Me&07,Pe&98}.
%(see, for instance, a paper of ours~\cite{De&17b}
%and the references therein).
The predicate $\textit{su\/}(M\!,R,\textit{Sum})$, which holds 
iff $\textit{Sum}\! =\! R\!+\!\sum_{x=1}^{M} x$, encodes the operational
semantics of the program \texttt{sum\_upto}. 
Clause~1 encodes the Hoare triple,  stating that if 
$\textit{su\/}(M\!,R,\textit{Sum})$
holds with $ M\!\geq\! 0$ (that is, the precondition ${\tt m}\!\geq\!0$ holds) and 
$R\!=\!0$ (that is, ${\tt r}$ is initialized to $0$), 
then $\textit{Sum}\!\geq\! M$ (that is, at the end of the execution,
the value of the variable ${\tt sum}$ 
is not smaller than the value of the variable~${\tt m}$). 
Clauses~2 and~3 encode the while-loop of the function~{\tt f}.

Constrained Horn clauses are syntactically the same as {\it constraint logic programs}~\cite{JaM94}. 
However, the term `constrained Horn clauses' is mostly used in the field of
program verification and, unlike `constraint logic programs', it is not 
associated with any operational meaning.
Moreover, most of the research on constrained Horn clauses is devoted to 
finding a model, expressible in the constraint theory, that proves the {\it satisfiability} of the clauses, whereas the operational semantics of
constraint logic programs is based on a refutation procedure that
looks for a proof of the {\it unsatisfiability} of the clauses.
In this respect, the techniques used for finding models
of constrained Horn clauses are closer to the ones proposed 
for the static analysis of constraint logic programs based on abstract interpretation~\cite{CoC77,BeK96}, where the
objective is to find an over-approximation of the least model of the program.
%
%Finally, the constraint theories developed in the verification techniques based
%on constrained Horn clauses are those of interest for modeling 
%properties of data, such as integer numbers, bit-vectors, pointers, and arrays~\cite{DeB08},
%which are have not received much attention in the field of 
%constraint logic programming.

The proof of satisfiability
of sets of constrained Horn clauses is supported by 
{\it CHC solvers} that have been developed in recent years
for various 
%%In recent years many {\rm CHC solvers} have been developed to verify 
%%the satisfiability of sets of constrained Horn clauses modulo various
constraint theories, such as
(linear or nonlinear) integer arithmetic, real (or rational) arithmetic, 
booleans, integer arrays, lists, heaps, and other data 
structures~\cite{De&14b,Gr&12,Gu&15,Ho&11,Ho&12,Ka&16,McR13}. 
%
%The use of solvers for proving satisfiability of clauses is of great importance
%because in many cases other techniques, such as those based on
%evaluations of constraint goals (with or without tabling), do not work. 
%%
%%For instance, 
%%for clauses 1--3 above, the evaluation of the goal consisting of clause~1
%%by using standard constraint logic programming systems,
%%will not terminate. Neither the tabling technique may be of help here, as infinitely many answers should be memoized.
%%In contrast, CHC solvers will prove satisfiability
%%by making use of constraint-based abstraction techniques.
%%
However, in general, 
since the satisfiability of constrained Horn clauses is an undecidable
problem, CHC solvers may not be able to return conclusive answers. 

In order to improve the effectiveness of CHC solvers, several techniques proposed
by recent papers perform satisfiability preserving 
transformations on sets of clauses that, in some cases, derive clauses whose satisfiability is easier to prove~\cite{De&14c,De&15c,De&15d,De&16c,KaG15a,KaG17}.
These transformations are adaptations to the task of improving the
effectiveness of satisfiability checking of earlier techniques 
which were developed for improving
the efficiency of execution of (constraint) logic programs, such as 
{\it query answer transformation}, {\it specialization} 
(or {\it partial deduction}), and 
{\it unfold/fold transformations}~\cite{DeR94,EtG96,LeB02,TaS84,PeP94}.

In this paper we further enhance the approach
to CHC satisfiability checking based on unfold/fold transformations.
Our two main contributions are the following: 
(1) we prove in a precise mathematical sense that the 
application of the unfold/fold transformation rules 
cannot worsen the effectiveness of 
CHC solvers, and actually these rules are able to strictly
enlarge the set of satisfiability problems
that can be solved by a given class of CHC solvers; and
(2) we provide a specific strategy, called {\it Predicate Pairing}, for applying the transformation rules with the objective of improving the ability
of CHC solvers to prove satisfiability.

The basic idea behind the first contribution is as follows.
Similarly to what is introduced in a paper by Bj{\o}rner et al.~\cite{Bj&15},
we consider the notion of the $\mathcal A$-{\it definable} model, which
is a model definable in a class $\mathcal A$ of first order formulas.
Typically, CHC solvers  (and, in particular, the solvers
based on {\it predicate abstraction})
look for models in specific classes, such
as {\it linear {\rm (}integer or real{\rm\,)} arithmetic} formulas, or
{\it quantifier-free  array} formulas.
%Then the power of a CHC solver can be characterized in terms of its
%{\it completeness} relative to $\mathcal A$. 
%A  solver is complete relatively to $\mathcal A$ if it founds an
%$\mathcal A$-model iff such a model exists.
While satisfiability is undecidable and not semidecidable, the
existence of an \mbox{$\mathcal A$-definable} model is semidecidable, as long as
the validity problem for the formulas in $\mathcal A$ is decidable, 
and hence solvers that find an $\mathcal A$-definable model whenever it exists,
can indeed be constructed.
We prove that, under very general conditions on $\mathcal A$,
the unfold/fold rules preserve the existence of an $\mathcal A$-definable 
model, that is,
if the original clauses have an $\mathcal A$-definable model, 
then also the transformed clauses have an $\mathcal A$-definable model.
The converse does not hold: there are cases where the 
original clauses have no $\mathcal A$-definable model, 
while the transformed clauses have an $\mathcal A$-definable model.
{In this sense the application of the unfold/fold rules
may improve the effectiveness of a CHC solver that works by searching for
$\mathcal A$-definable models, because the solver
may be able to find an $\mathcal A$-definable model after the transformation 
in cases where there was no such a model before the transformation.}

We also provide less general conditions under which 
the transformed clauses have an $\mathcal A$-definable model {\it if and only if}
the original ones have an $\mathcal A$-definable model. These conditions
prevent the introduction of new predicates
that have recursive definitions
in terms of the old predicates occurring in the original clauses.
Thus, the source of the improvement of the effectiveness
of the CHC solver due to the \mbox{unfold/fold} transformations
is the introduction of one or more new predicates and 
the derivation of new (mutually) recursive definitions for these predicates.

The second contribution of our paper is related to the fact that, 
due to the already mentioned undecidability limitations, there is no
universal algorithm that, starting from a set of clauses,
applies the unfold/fold rules and derives a set of clauses such that, 
if it is satisfiable, then it
has an $\mathcal A$-definable model, for some theory~$\mathcal A$ whose
validity problem is decidable.
%, whenever it is satisfiable
%, then they have an $\mathcal A$-definable model 
%{(assuming that the validity 
%problem for formulas in $\mathcal A$ is decidable).}
Therefore, it should not be unexpected 
that the {\rm Predicate Pairing} strategy we propose for guiding the
use of the unfold/fold transformation rules is 
based on heuristics. We show that this strategy is capable, 
in many significant cases, of
transforming sets of clauses into new, equisatisfiable sets of clauses,
whose satisfiability problem can be solved by constructing
$\mathcal A$-definable models, while the original sets 
have no $\mathcal A$-definable models.
Predicate Pairing introduces a new predicate defined by the conjunction
of two predicates together with a conjunction of constraints. 
We will explain through examples why an $\mathcal A$-definable model
may exist for a conjunction of predicates, even if it does not
exist for the atomic conjuncts in isolation. (Obviously, 
by a repeated application of Predicate Pairing, we may introduce new predicates
corresponding to the conjunction of more than two old predicates.)
Thus, Predicate Pairing can be viewed as an extension to constrained
Horn clauses of techniques for transforming logic programs,
such as the {\it tupling} unfold/fold strategy~\cite{PeP94} and {\it conjunctive partial deduction}~\cite{De&99}. 

We will show that constraint-based reasoning is essential
for guiding the introduction of the suitable pairs of predicates
during the transformation process. Moreover, we will show that
Predicate Pairing works well for solving many satisfiability problems  
that arise from the field of imperative program verification. 
%, when using constrained Horn clauses. 
In particular, Predicate Pairing is a crucial technique for verifying 
{\it relational program properties}~\cite{Ba&11}, that is, 
properties relating two programs (such as program
equivalence) or two executions of the same program (such as non-interference).

The paper is structured as follows.
In Section~\ref{sec:prelim} we recall the basic notions concerning 
constrained Horn clauses and we define the notion of an $\mathcal A$-definable model.
In Section~\ref{sec:rules} we prove our results concerning the preservation of 
$\mathcal A$-definable models when using the unfold/fold transformation rules.
In Section~\ref{sec:strategy} we present the Predicate Pairing strategy and
in Section~\ref{sec:case_studies} we show some examples of its application
for verifying relational program properties. 
In Section~\ref{sec:experiments} we report the results obtained
by our implementation of that strategy by using the VeriMAP
transformation system~\cite{De&14b}.
Finally, in Section~\ref{sec:related} we discuss related work in the
fields of program transformation and verification.

\section{Constrained Horn Clauses}
\label{sec:prelim}
In this section we recall the basic definitions concerning 
constrained Horn clauses and their satisfiability, and we introduce the notion of
an $\mathcal A$-{\it definable model}.

\medskip

Let $\mathcal L$ be a first order language with equality
and $\textit{Pred}_u\subseteq \mathcal L$ be a set of predicate
symbols, called the {\it user-defined} predicate symbols.
Let $\mathcal C$  be a set 
of formulas of $\mathcal L \setminus \textit{Pred}_u$, called the set of 
 {\it constraints}.
We assume that: (i)~$\textit{true}$, $\textit{false}$, and
equalities between terms belong to~$\mathcal C$,
and (ii)~$\mathcal C$ is closed under conjunction. 

An {\it atom} is an atomic formula of the form $p(X_{1},\ldots,X_{m})$,
where $p$ is a predicate symbol in $\textit{Pred}_u$ and 
$X_{1},\ldots,X_{m}$ are distinct variables.
Let {\it Atom} be the set of all atoms.
A~{\it definite constrained clause} is an implication of the form  
%$c \wedge A_1 \wedge \ldots \wedge A_n\rightarrow A$, 
\mbox{$c \wedge G\rightarrow H$} whose
premise (or {\it body\/}) is the conjunction of a constraint $c$ 
%and~$A_1 \wedge \ldots \wedge A_n$ 
and~a (possibly empty)
conjunction $G$ of $n\, (\geq 0)$ atoms $A_1 \wedge \ldots \wedge A_n$, and 
whose conclusion (or {\it head\/}) $H$ is an atom. 
A~{\it constrained goal} (or simply, a {\it goal}\/) is  
an implication of the form  
$c \wedge G \rightarrow \textit{false}$, where $c$ is a constraint and $G$ is a conjunction of atoms.
A~{\it constrained Horn clause} (CHC) (or simply, a {\it clause}) 
is either a definite constrained
clause or a constrained goal. 
A~set of constrained Horn clauses is said to be a {\it CHC set}.
A~constrained Horn clause $\textit{Cl}$ (or a set~$P$ of clauses) 
is said to be `{\it over~$\mathcal C$\,}' 
in case we want to stress that the constraints occurring in clause~$\textit{Cl}$ 
(or in the set~$P$ of clauses) are 
taken from the set $\mathcal C$ of constraints.
A~clause $c\wedge G \rightarrow H$
is said to be {\it linear} if $G$ consists of at most one atom, and {\it nonlinear} otherwise.

We will often use the logic programming syntax
and we write $H\leftarrow c,A_1, \ldots, A_n$, instead of $c \wedge A_1 \wedge \ldots \wedge A_n\rightarrow H$.
We will also feel free to write non-variable terms as arguments 
of atoms. Thus, the clause  
$p(\ldots,t,\ldots) \leftarrow c, G$ should be viewed as a shorthand for 
$p(\ldots,X,\ldots) \leftarrow X\!=\!t, c, G$, where $X$ 
is a variable not occurring elsewhere in the clause, and likewise,
$H \leftarrow c, G_1, p(\ldots,t,\ldots), G_2$ should be viewed as a shorthand for 
$H \leftarrow X\!=\!t, c, G_1, p(\ldots,X,\ldots), G_2$.

Given a formula $\varphi\!\in\!{\mathcal L}$, we denote by $\exists (\varphi)$
its {\it existential closure} and by $\forall (\varphi)$ its 
{\it universal closure}. By $\textit{vars}(\varphi)$ and 
$\textit{Fvars}(\varphi)$ we denote the set of variables and the set of 
the free variables,
respectively, occurring in $\varphi$.

For the notions of an {\it interpretation} and a {\it model} of a first order formula
we will use the standard notions and notations~\cite{Men97}.
We fix a {\it canonical interpretation}~$\mathbb D$ of the symbols in
$\mathcal L \setminus \textit{Pred}_u$.
A {\it ${\mathbb D}$-interpretation} is an interpretation of~$\mathcal L$
that for all symbols occurring in~$\mathcal L \setminus \textit{Pred}_u$, agrees with
${\mathbb D}$.
If $U$ is the universe of~${\mathbb D}$, then
a  ${\mathbb D}$-interpretation $\mathbb I$ can be identified with the  set
of atoms:

$\{p(a_1,\ldots,a_m) \ | \ (a_1,\ldots,a_m) \in U^m \mbox{ and }
p^{\mathbb I}(a_1,\ldots,a_m) \mbox{ holds in } \mathbb I\}$

\noindent
where $p^{\mathbb I}$ denotes the $m$-ary relation which is
the interpretation of $p$ in $\mathbb I$.
Given any set $F$ of formulas, a ${\mathbb D}$-interpretation
$\mathbb M$ is a {\it ${\mathbb D}$-model} of $F$, written
$\mathbb M \models  F$, if, for all formulas 
$\varphi\!\in\! F$, $\mathbb M \models \varphi$ holds. 
$ F$ is ${\mathbb D}$-{\it satisfiable} if
it has a ${\mathbb D}$-model.
{We will feel free to say {\it satisfiable}, instead of 
${\mathbb D}$-{\it satisfiable}, when the interpretation ${\mathbb D}$
is clear from the context.}	
We write ${\mathbb D} \models  F$
if, for every ${\mathbb D}$-interpretation $\mathbb M$, $\mathbb M \models F$
holds.

%We say that a set $\mathcal S$ of clauses is ${\mathbb D}$-{satisfiable} if
%the set $\{\forall(C) ~|~ C\in \mathcal S\}$ is ${\mathbb D}$-{satisfiable}.
A set $P$ of definite constrained  clauses is ${\mathbb D}$-{satisfiable}
and has a {\it least} (with respect to 
set inclusion) ${\mathbb D}$-model,
denoted $\textit{lm}(P)$~\cite{JaM94}.
Thus, if $P$ is any set of constrained Horn clauses and
$Q$ is the subset of the constrained goals in~$P$, then
$P$ is ${\mathbb D}$-{satisfiable}
if and only if $\textit{lm}(P \setminus Q)\models Q$.

%Thus, if  $\mathcal G$ is a set of constrained goals, 
%$\mathcal S \cup \mathcal G$ is ${\mathbb D}$-{satisfiable}
%if and only if $\textit{lm}(\mathcal S)\models \mathcal G$.

Many CHC solvers based on {\it predicate abstraction}~\cite{Bj&15}
try to check the ${\mathbb D}$-satisfiability 
of a set of constrained Horn clauses 
by looking for the existence of ${\mathbb D}$-models that are definable
by formulas belonging to a given set $\mathcal A$, which is a subset of the 
set $\mathcal C$ of constraints. 
This restriction when looking for models may significantly
ease the satisfiability test, as shown by the following example.

\begin{example}
	\label{ex:2VAR}
Let us assume that $\mathcal C$ is the set of {\it linear integer arithmetic} (LIA)
constraints, that is, equalities $(=)$ and inequalities $(>)$ between linear polynomials
with integer coefficients and integer-valued variables, closed with respect to conjunction
and disjunction. 
We also use the symbols $\geq$, $\leq$, $<$, and $\neq$ with the usual definitions in terms of~$=$
and~$>$.
Let $\mathbb Z$ denote
the usual interpretation of integer arithmetic.

Now, let us consider clauses 1--3 listed in the Introduction.
The satisfiability of these clauses can be proved by looking for
models that are definable by constraints~$\varphi$ in the subset of LIA, which we call 2VAR,  
defined by the following grammar:

\vspace{1mm}
$\varphi ::= \textit{true}  \ | \ \textit{false}  \ | \  X\!>\!0 \ | \  X\!=\!0 \ | \  X\!>\!Y \ | \  X\!=\!Y \ | \ \varphi_1 \wedge \varphi_2\ | \ \varphi_1 \vee \varphi_2$

\vspace{1mm}
\noindent
where $X$ and $Y$ are variables. 
Thus, 2VAR is the set of linear integer constraints constructed from 
arithmetic comparisons between at most two variables,
and it is  a subset of the {\it Octagons} domain often considered in
the field of abstract interpretation~\cite{Min06}.
A 2VAR-definable model of clauses 1--3 is given
by interpreting the predicate 
$\textit{su\/}(M\!,R,\textit{Sum})$
as the set of triples satisfying the following constraint in 2VAR:

$(\textit{Sum}\!\geq\! M \wedge \textit{Sum}\!\geq\! R) \ \vee\ R\!< \!0$,

\noindent
In order to show that the above interpretation indeed defines a 
$\mathbb Z$-model, 
we replace the instances of  $\textit{su}(M,R,\textit{Sum})$ by 
the corresponding instances of the formula
\mbox{$(\textit{Sum}\!\geq\! M$ $\wedge$} $\textit{Sum}\!\geq\! R) \vee R\!< \!0$ in
clauses~1--3, and we check that
the resulting implications hold in $\mathbb Z$. \hfill$\Box$
\end{example}

This Example~\ref{ex:2VAR} motivates the introduction of the  notion of an 
$\mathcal A$-definable model, which is a generalization
of the one presented in the literature~\cite{Bj&15}, 
where $\mathcal A$ coincides with~$\mathcal C$.

\begin{definition}\label{def:SymbInt}
Let $\mathcal A\!\subseteq\! \mathcal C$ be a set of formulas of 
$\mathcal L \setminus \textit{Pred}_u$ such that: (i)~$\textit{true}$, 
$\textit{false}$, and equalities between terms belong to~$\mathcal A$,
and (ii)~$\mathcal A$~is closed under conjunction.
{Let $\mathbb D$ be the {canonical interpretation} of the symbols in
$\mathcal L \setminus \textit{Pred}_u$.}
We denote by $\mathcal A^{\exists \vee}$ the set of formulas 
$\{\exists X_1\! \ldots \exists X_m (\varphi_1 \vee \ldots \vee \varphi_n) \ | \  m\!\geq\! 0, n\!>\! 0,$ and for $i\!=\!1,\ldots,n,\ \varphi_i\! \in\! 
\mathcal A \}$.
A {\it symbolic interpretation} is a function $\Sigma \!: \textit{Atom}   \cup \{\textit{false}\}
\longrightarrow \! \mathcal A^{\exists \vee}$ such that  $\Sigma(\textit{false})=\textit{false}$ and, 
for every \mbox{$A\! \in\!\textit{Atom}$}, 
(i)~$\textit{Fvars}(\Sigma(A))\subseteq \textit{Fvars}(A)$, and
(ii)~for every renaming substitution $\rho$ for $A$~\cite{Llo87}, 
$\Sigma(A\rho) = \Sigma(A)\rho$. 
We extend $\Sigma$ to conjunctions of atoms by stating that $\Sigma(A_1\wedge\ldots\wedge A_n) = \Sigma(A_1) \wedge\ldots\wedge \Sigma(A_n)$.
Given a set $P$ of constrained Horn clauses over $\mathcal C$, 
a symbolic interpretation $\Sigma$ is an 
$\mathcal A$-{definable model} of $P$, written $\Sigma\models P$, if  for every clause $H\! \leftarrow\! c, A_1,\!\ldots,\! A_n$ in $P$, 
$\mathbb D\models\forall(c \wedge \Sigma(A_1\!\wedge\ldots\wedge\! A_n) \rightarrow \Sigma(H))$ holds.
\end{definition}

Note that the symbolic interpretation $\Sigma$ of
an atom is independent of the variable names occurring in that atom, 
and hence, for each predicate symbol $p$, the formula 
$\Sigma(p(X_1, \ldots, X_m))$ is unique up to variable renaming.
Note also that the definition of a symbolic interpretation is essentially equivalent 
to that given by Kafle and Gallagher, who define an interpretation as a set of 
{\it constrained facts} of the form $p(X_1, \ldots, X_m) \leftarrow c$, where
$c$ is a constraint in $\mathcal A$~\cite{KaG17}. Indeed,
the set of constraints in the bodies of the constrained facts with the same head
predicate can be represented as a disjunction of those constraints,
and the variables occurring in the body of a constrained fact and not in its head
are implicitly existentially quantified.

Clearly, if $P$ has an $\mathcal A$-definable model, then
$P$ is ${\mathbb D}$-satisfiable. 
In general the converse does not hold, as shown by the following 
example.

\begin{example}
\label{ex:notAsolvable}

%\comment{REVIEWER:
%p.7 Example 2: I ike the example, but please say  explicit what are the notions of Def. 1 in this example: 
%what is $\cal C$ , what is $\cal A$ , what is the function $\sigma$ for su/3 and sq/4, .... }
%

Let us continue Example~\ref{ex:2VAR}, where the sets of constraints
$\mathcal{C}$ and $\mathcal{A}$ are LIA and 2VAR, respectively.
Let us consider the program in Figure~\ref{fig:square}, which computes
the square of a non-negative integer {\tt n} by summing up {\tt n}
times the value of {\tt n}.   

\begin{figure}[h]
\vspace{-1mm}
\centering
\begin{minipage}[t]{0.31\textwidth}
\begin{verbatim}
int n, sqr;
int g(int y, int k) {
  int s = 0;
  while (y > 0) {
    s = s + k;   y--; } 
  return s; 
  }
\end{verbatim}  
\end{minipage}

\vspace{2mm}
%\vline ~ 
\begin{minipage}[t]{0.31\textwidth}
\begin{verbatim}
void square() {
  sqr = g(n,n); 
  }
\end{verbatim}       
\end{minipage}
\caption{\label{fig:square} Program {\tt square} computing 
${\mathtt{sqr}} ={\mathtt n}^{2}$.}
\vspace{-2mm}
\end{figure}

\noindent
Clauses 4--6 below express the following
property, which relates program {\tt sum\_upto}
and program {\tt square}: if the value of {\tt m} is equal to the value 
of {\tt n} before
the execution of the programs {\tt sum\_upto} and {\tt square} and they 
both terminate,
then at the end of their execution the value of {\tt sqr} is not smaller than
the value of {\tt sum}.
%, that is, for any ${\mathtt m}\!\geq\! 0$, we have that
%${\mathtt{m}}^{2} \geq \sum_{i=0}^{\mathtt m} $.
Note that, since the programs {\tt sum\_upto} and {\tt square} have disjoint
sets of variables, the order of their execution is immaterial.

\vspace{1mm}
\noindent
4. \ $\textit{false} \leftarrow \textit{Sum}\!>\!\textit{Sqr},\  M\!\geq\!0,\ M\!=\!N,\ N\!=\!Y,\ 
R0\!=\!0,\ S0\!=\!0,\ $

\hspace{14mm}$\textit{su\/}(M\!,R0,\textit{Sum}),\ \textit{sq\/}(N\!,\!Y\!,S0,\textit{Sqr})$

\noindent
5. \ $\textit{sq\/}(K\!,\!Y\!,S0,S) \leftarrow Y\!\leq\!0,\ S\!=\!S0$

\noindent
6. \ $\textit{sq\/}(K\!,\!Y\!,S0,S) \leftarrow Y\!>\!0,\ Y1\!=\!Y\!-\!1,\ S1\!=\!S0\!+\!K,\ 
\textit{sq\/}(K\!,\!Y1,S1,S)$
\end{example}

\noindent
For $Y\!\geq\!0$, the atom $\textit{sq\/}(K\!,\!Y\!,S0,S)$  holds iff $S\!=\!S0+(K\!\!\times\!\!Y)$.
Properties like the one between programs {\tt sum\_upto} and {\tt square}
are called {\it relational properties} \cite{Ba&11}. 
Similarly to the verification conditions for partial correctness properties, 
the clauses  for relational properties, also called verification conditions,
can be automatically generated  
from the formal specification of those properties and the operational semantics
of the programming language~\cite{De&16c}. For brevity, we do not
give here the details of that generation process, which is inessential for understanding
the techniques presented in this paper.

Clauses 2--6 are constrained Horn clauses over LIA and they
are $\mathbb Z$-satisfiable. 
Indeed, in the least $\mathbb Z$-model of clauses 2, 3, 5, 6,
for all integers $M,\textit{Sum},$ and $\textit{Sqr},$ with $M\!\geq\!0$,
if $\textit{su}(M,0,\textit{Sum})$ and $\textit{sq\/}(M,M,0,\textit{Sqr})$ hold,
then $\textit{Sum}\!\leq\!\textit{Sqr}$ holds.
However, clauses  2--6 do not admit a 2VAR-definable model.
Indeed, no constraint of the form $\textit{Sum}\!\leq\! X$, for any variable $X$,
is a consequence of $\textit{su}(M,0,\textit{Sum})$, and hence we cannot
infer $\textit{Sum}\!\leq\!\textit{Sqr}$, independently of the
constraints that are consequences of $\textit{sq\/}(M,M,0,\textit{Sqr})$.
Actually, it is not difficult to see that a similar limitation holds even if we look for a
LIA-definable model, rather that a 2VAR-definable model. Indeed, in order to infer
the constraint $\textit{Sum}\!\leq\!\textit{Sqr}$ 
one should discover quadratic relations, such as
$\textit{Sum} \!=\! M\!\times\!(M\!-\!1)/2$ and 
$\textit{Sqr}\!=\!M\!\times\!M$, starting 
from $\textit{su}(M\!,0,\textit{Sum})$ and 
$\textit{sq\/}(M\!,M\!,0,\textit{Sqr})$,
respectively, and these relations cannot be expressed by linear arithmetic constraints.\hfill$\Box$

\section{Transformation Rules and Preservation of $\mathcal A$-definable Models}
\label{sec:rules}
%\comment{REVIEWER: p.8 first line of section 3 mentions $\cal C$ while in R1 you have $\cal A$: error?}

Let $\mathcal C$ be a set of constraints and $\mathcal A \subseteq \mathcal C$.
A {\it transformation sequence over $\mathcal C$} 
is a sequence of CHC sets 
$P_0,P_1,\ldots,P_n$ over $\mathcal C$,
where, for $i\!=\!0,\ldots,n\!-\!1,$ $P_{i+1}$ is derived from $P_i$ by
applying one of the following \mbox{rules~R1--R4}.

Let $\textit{Defs}_i$ denote the set of all the clauses, 
called {\it definitions}, 
introduced by rule~R1 during the construction of the transformation sequence
$P_0,P_1,\ldots,P_i$. 
Thus, $\textit{Defs}_0\!=\!\emptyset$.

\medskip
\noindent
(R1)~{\it Definition.}  
We introduce a clause $D$: $\textit{newp}(X_1,\ldots,X_k)\leftarrow c,G$, 
where:
(i)~\textit{newp} is {a predicate symbol in $\textit{Pred}_u$ not occurring in 
the sequence $P_0,P_1,\ldots,P_i$,}
\mbox{(ii)~$c \!\in\! \mathcal A$,} 
(iii)~$G$ is a non-empty conjunction of atoms whose predicate symbols 
occur in $P_0$, and 
(iv)~$X_1,\ldots,X_k$ are distinct variables occurring free in $(c,G)$.
Then, we derive the new set $P_{i+1}=P_i\cup \{D\}$ and
$\textit{Defs}_{i+1}=\textit{Defs}_i \cup \{D\}$.

\medskip
\noindent
(R2)~{\it Unfolding.} 
Let  $C$: $H\leftarrow c,G_1,p(X_1,\ldots,X_k),G_2$ be a clause in $P_i$.
Let\nopagebreak 

$\{p(X_1,\ldots,X_k)\leftarrow {c}_j, B_j ~\mid~  j\!=\!1, \ldots, m\}$ 

\noindent
be the (possibly empty) set of clauses in~$P_i$ whose head predicate is $p$. 
Without loss of generality, we assume that, for $j=1, \ldots, m,$ 
$\textit{vars}({c}_j, B_j)\cap\textit{vars}(C) \subseteq \{X_1,\ldots,X_k\}$.
By {\it unfolding the atom $p(X_1,\ldots,X_k)$ in $C$ using $P_i$}
we derive the new set~$P_{i+1}=(P_i\setminus\{C\}) \cup 
\{H\leftarrow  c, {c}_j,G_1, B_j, G_2 \mid  j=1, \ldots, m\}$.

\medskip
\noindent
(R3)~{\it Folding.} 
Let $C$: $H\leftarrow c, G_1,Q,G_2$ be a clause in $P_i$, where 
$Q$ is a non-empty conjunction of atoms, and let
$D$: $K \leftarrow d, B$ be (a variant of) a clause in $\textit{Defs}_i$
with $\textit{vars}(C)\cap \textit{vars}(D)\!=\!\emptyset$.
Suppose that there exist a substitution~$\vartheta$ 
and a constraint $e$ such that:
(i)~\mbox{$Q\!=\! B\vartheta $,}
(ii)~$\mathbb D\models \forall(c \leftrightarrow (e\! \wedge\! d\vartheta))$, 
%(ii)~$\mathbb D\models \forall(c \rightarrow d\vartheta)$, 
and
(iii)~for every variable \( X\!\in\!\textit{vars}(d,B)\setminus\textit{vars}(K)\),
the following conditions hold: (iii.1) \(
X\vartheta \) is a variable not occurring in \( \{H,c,G_{1},G_{2}\}
\), and (iii.2)~\( X\vartheta  \) does not occur in the term \(
Y\vartheta  \), for any variable \( Y \) occurring in \( (d, B) \)
and different from \( X \).
By \textit{folding \( C\)
using the definition \( D\)}, we derive clause 
\(E  \):~\( H\leftarrow e, G_1, K\vartheta, G_{2} \).
In this case we also say that $E$ is derived 
\textit{by folding $Q$ in $C$}.
%\(E  \):~\( H\leftarrow c, G_1, K\vartheta, G_{2} \).
We derive the new set \( P_{i+1}=(P_{i}\setminus\{C\})\cup \{E \} \).

\medskip
\noindent
(R4)~{\it Constraint Replacement.} 
Let us consider a subset of $P_i$ of the form \linebreak  %% for better layout +++
$\{(H\leftarrow c_1, G), \ldots,(H\leftarrow c_k, G)\}$. 
Suppose that, for some constraints $d_1,\ldots,d_m,$ 

$\mathbb D \models \forall \, (\exists Y_1\ldots\exists Y_r\ (c_1 \vee \ldots \vee c_k) \leftrightarrow 
\exists Z_1\ldots\exists Z_s\ (d_1 \vee \ldots \vee d_m))$

\noindent
where $\{Y_1,\ldots,\!Y_r\}\!=\!Fvars(c_1 \vee \ldots \vee c_k)\setminus vars(\{H,\!G\})$
and $\{Z_1,\ldots,\!Z_s\}\!=\!Fvars(d_1 \vee \ldots \vee d_m)\setminus vars(\{H,G\})$.
Then, we derive the new set 
\noindent
$P_{i+1} = (P_i \setminus\{(H\leftarrow c_1, G), \ldots,$ $(H\leftarrow c_k, G)\})$ 
$ \cup $ $\{(H\leftarrow d_1, G), \ldots,(H\leftarrow d_m, G)\}$.

\medskip

Note that rule R4 enables the deletion of a clause with an inconsistent constraint 
in its body. Indeed, if $c_1$ is unsatisfiable, then $\mathbb D \models \forall \, (c_1 \leftrightarrow d_1 \vee \ldots \vee d_m)$ 
with~$m\!=\!0$.

The following result~\cite{EtG96} shows that the transformation rules R1--R4
derive sets of clauses that are equivalent with respect to the least $\mathbb D$-model.

\noindent
\begin{theorem}[Equivalence with respect to the Least $\mathbb D$-Model]
\label{thm:leastmodel}
Let $P_0,P_1,\ldots,P_n$ be a transformation sequence where,
for $i=0,\ldots,n,$ $P_i$ is a set of definite clauses.
Let us assume that every definition in $\textit{Defs}_n$ is unfolded during the
construction of this sequence (that is, for every definition $D\!\in\!\textit{Defs}_n$, 
there exists~$i$, with $0\!\leq\! i\!\leq\! n\!-\!1$, such that $P_{i+1}$ is derived from 
$P_{i}$ by unfolding $D$).
Then, for every predicate~$p$ and $(a_1,\ldots,a_m) \in U^m$,

\vspace{.5mm}

$p(a_1,\ldots,a_m)\! \in \!\textit{lm}(P_0\cup \textit{Defs}_n)$ ~if and only if~ $p(a_1,\ldots,a_m)\! \in\! \textit{lm}({P_n})$.
\end{theorem}

\noindent
From Theorem~\ref{thm:leastmodel} it follows that, as we now show, the
transformation rules R1--R4 derive sets of clauses that are equivalent
with respect to $\mathbb D$-satisfiability.

\begin{theorem}[Equivalence with respect to $\mathbb D$-Satisfiability]
\label{thm:D-sat}
Let $P_0,P_1,\ldots,P_n$ be a transformation sequence
such that every definition in $\textit{Defs}_n$ is unfolded during the 
construction of this sequence.
Then, $P_0$ is $\mathbb D$-satisfiable if and only if $P_n$
is $\mathbb D$-satisfiable.
\end{theorem}

\vspace{-1mm}
\begin{proof}
First we observe that $P_0$ is $\mathbb D$-satisfiable iff 
$P_0 \cup \textit{Defs}_n$ is $\mathbb D$-satisfiable. Indeed: 
(i)~if~$M$ is a $\mathbb D$-model of $P_0$, then the $\mathbb D$-interpretation
$M \cup \{\textit{newp}(a_1,\ldots,a_k) ~\mid~ \textit{newp}$ 
is a head predicate
in $\textit{Defs}_n \mbox{ and } (a_1,\ldots, a_k) \in U^k\}$ 
is a $\mathbb D$-model of $P_0 \cup \textit{Defs}_n$,
and (ii)~if $M$ is a $\mathbb D$-model of $P_0 \cup \textit{Defs}_n$, then
by the definition of $\mathbb D$-model,
$M$  is a $\mathbb D$-model of $P_0$.

Now let us consider a new sequence  $P'_0,P'_1,\ldots,P'_n$
obtained from the transformation sequence $P_0,P_1,\ldots,P_n$
by replacing each occurrence of \textit{false} in the head of
a clause by a new predicate
symbol $f$.
The sequence $P'_0,P'_1,\ldots,P'_n$ satisfies the 
hypothesis of Theorem~\ref{thm:leastmodel}, and hence 
$f\! \in \!\textit{lm}(P'_0\cup \textit{Defs}_n)$ iff 
$f\! \in\! \textit{lm}({P'_n})$. 

We have that:
$P_0 \cup \textit{Defs}_n$ is $\mathbb D$-satisfiable
iff $P'_0 \cup \textit{Defs}_n\cup \{\neg f\}$ is $\mathbb D$-satisfiable 
iff $f\not\in\textit{lm}(P'_0 \cup \textit{Defs}_n)$ 
iff \{by Theorem~\ref{thm:leastmodel}\} $f\not\in\textit{lm}(P'_n)$ 
iff $P'_n\cup \{\neg f\}$ is $\mathbb D$-satisfiable
iff $P_n$ is $\mathbb D$-satisfiable.\hfill
%%
%%\makebox[41mm][l]{} iff $f\not\in\textit{lm}(P'_0 \cup \textit{Defs}_n)$
%%
%%\makebox[41mm][l]{} iff $f\not\in\textit{lm}(P'_n)$ 
%%(by Theorem~\ref{thm:leastmodel})
%%
%%\makebox[41mm][l]{} iff $P'_n\cup \{\neg f\}$ is $\mathbb D$-satisfiable
%%
%%\makebox[41mm][l]{} iff $P_n$ is $\mathbb D$-satisfiable.\hfill
\end{proof}

Theorem~\ref{thm:D-sat} is {\it not} sufficient to ensure that a 
transformation sequence
preserves the existence of an $\mathcal A$-definable model.
Indeed, as shown by Example~\ref{ex:notAsolvable} in 
Section~\ref{sec:prelim}, for some
set $\mathcal A$ of constraints, $\mathbb D$-satisfiability
does not imply the existence of an $\mathcal A$-definable model.

Now, in order to study the preservation of $\mathcal A$-models during 
the construction of a transformation sequence over $\mathcal C$,
for $\mathcal A\!\subseteq\! \mathcal C$, 
we introduce the notions of {\it $\mathcal A$-soundness} and 
{\it $\mathcal A$-completeness}.

\vspace{-1mm}
\begin{definition}[$\mathcal A$-Soundness, $\mathcal A$-Completeness]
Let $P_0,P_1,\ldots,P_n$ be a transformation sequence.
(i)~If $P_0$ has an $\mathcal A$-definable model implies that $P_n$ has an $\mathcal A$-definable model, we say that the sequence is {\it $\mathcal A$-sound}.
(ii)~If $P_n$ has an $\mathcal A$-definable model implies that $P_0$ has an $\mathcal A$-definable model, we say that the sequence is \mbox{\it $\mathcal A$-complete}.
\end{definition}
\vspace{-1mm}

In order to prove the $\mathcal A$-soundness of a transformation sequence (see
Theorem \ref{thm:A-sound} below) we need the following 
definition and theorem.

\vspace{-1mm}
\begin{definition}{\label{def:tight-interpretation}}
Let $\mathcal S$ be a CHC set. A symbolic interpretation $\Sigma$ is said to be
{\it tight on}  $\mathcal S$ if for all clauses
$A\leftarrow c,G$ in $\mathcal S$, 
$\mathbb D\models \forall( \Sigma (A) \leftrightarrow \exists X_1\ldots\exists X_k (c\wedge \Sigma (G)))$, where $\{X_1, \ldots, X_k\}= \textit{Fvars}(c\wedge \Sigma(G)) \setminus \textit{Fvars}(\Sigma(A))$.
\end{definition}

{For instance, given the singleton set of clauses $\mathcal S=\{p(X) \leftarrow q(X)\}$,
the symbolic interpretation $\Sigma_1$ that maps both $p(X)$ and $q(X)$ to $X\!=\!0$ is tight on $\mathcal S$,
while the symbolic interpretation $\Sigma_2$ that maps $p(X)$ to \textit{true} and $q(X)$ to $X\!=\!0$
is not tight on $\mathcal S$. Both $\Sigma_1$ and $\Sigma_2$ are models.
}

\begin{theorem}
\label{thm:tight}
Let $P_0,P_1,\ldots,P_n$ be a transformation sequence.
For $i=0,\ldots,n-1,$ if $P_i$ has an $\mathcal A$-definable model that is
tight on $\textit{Defs}_i$, then $P_{i+1}$
has an $\mathcal A$-definable model that is tight on $\textit{Defs}_{i+1}$.
\end{theorem}

\vspace{-1mm}
\noindent {\it Proof} ~See Appendix. ~~~ $\Box$
%\begin{proof} See Appendix. \end{proof}

\smallskip

{The hypothesis that $P_i$ has an $\mathcal A$-definable model that is
tight on $\textit{Defs}_i$ is needed to guarantee that the folding rule
replaces a conjunction consisting of constraints and atoms by a single atom which is 
{\it equivalent} in the given model.}

\vspace{1mm}
From Theorem \ref{thm:tight} and the fact that,
if $P_0$ has an $\mathcal A$-definable model, then $P_0$ has an $\mathcal A$-definable model
that is tight on $\textit{Defs}_0$, which is the empty set,
we get the following result.

\vspace{-1mm}
\begin{theorem}[$\mathcal A$-Soundness]
\label{thm:A-sound}
Every transformation sequence is $\mathcal A$-sound.
\end{theorem}
\vspace{-1mm}

Now we prove that, if some suitable hypotheses hold, a
transformation sequence is also $\mathcal A$-complete. First, we need
the following two definitions.

\vspace{-1mm}
\begin{definition}\label{self-unfolding}
An application of the unfolding rule R2 to a clause 
 $C$: $H\leftarrow c,G_1,p(X_1,\ldots,X_k),G_2$ in $P_i$
is said to be a {\it self-unfolding} if the predicate of $H$ is $p$.
\end{definition}

\vspace{-2mm}
\begin{definition}\label{rev-folding}
Let $P_0,P_1,\ldots,P_n$ be a transformation sequence.
An application of the folding rule R3 to a clause 
$C$ in~$P_i$ using a definition $D$ in $\textit{Defs}_i$
is said to be a {\it reversible folding} if $D$ belongs to $P_i$ and is different from~$C$.
\end{definition}

\vspace{-2mm}
\begin{theorem}[$\mathcal A$-Completeness]
\label{thm:C-compl}
Let $P_0,P_1,\ldots,P_n$ be a transformation sequence over a set $\mathcal C$
of constraints. Let $\mathcal A$ be equal to $\mathcal C$.
Suppose that:
(i) no application of the unfolding rule is a self-unfolding, and 
(ii) every application of the folding rule is a reversible folding. 
Then, $P_0,P_1,\ldots,P_n$ is $\mathcal A$-complete.
\end{theorem}
\vspace{-1mm}

\noindent {\it Proof} ~See Appendix. ~~~ $\Box$
%\begin{proof} See Appendix. \end{proof}

%\comment{p.9-10  Explain better the relevance/role of the 2 notions (soundness and completeness) : 
%maybe add a picture illustrating your results. }

\medskip
In Section \ref{sec:strategy} we will present the Predicate Pairing transformation strategy
and we will show that it  generates transformation sequences that are $\mathcal A$-sound,
but not necessarily $\mathcal A$-complete (see Theorem~\ref{thm:soundness}).
	
Normally, $\mathcal A$-soundness is a desirable property of a transformation sequence\linebreak $P_0,P_1,\ldots,$ $P_n$.
Indeed, suppose we have a CHC solver, call it \textit{SOLVE}, that finds an 
\mbox{$\mathcal A$-definable} model
of a set of clauses whenever it exists. As already mentioned, such an ideal solver exists,
as long as the validity problem for the formulas in $\mathcal A$ is decidable.
Then, $\mathcal A$-soundness guarantees that if the satisfiability of $P_0$ can be
proved by using \textit{SOLVE}, then also the satisfiability of $P_n$ can be proved 
by using \textit{SOLVE}. In other terms, the effectiveness of the solver is not worsened by
the transformation.

In contrast, $\mathcal A$-completeness might not always be a desirable property. 
Indeed, in many cases we may want to transform clauses for which
\textit{SOLVE} cannot find an $\mathcal A$-definable model, because such a model does
not exist, and derive equisatisfiable clauses with an $\mathcal A$-definable model
which can be constructed by using \textit{SOLVE}.

In practice, the existing solvers do not guarantee that they find 
an $\mathcal A$-definable model of a set of clauses whenever it exists.
Thus, the theoretical properties of $\mathcal A$-soundness and $\mathcal A$-completeness
might not hold in some cases.
We will show through the experiments reported in Section~\ref{sec:experiments}, 
that these unfortunate cases are rare.

\vspace{1mm}
We conclude this section by showing that there are $\mathbb D$-safisfiable
CHC sets that have no $\mathcal A$-definable models, and yet can be transformed, by 
applying rules R1--R4, into  CHC sets that have $\mathcal A$-definable models.

\begin{example}\label{ex:notAsolvableCont}
%	\comment{MAU: Add $N\geq 0$?}
	
%\comment{REVIEWER: p. 10 Example 3, please link it up with the previous notions: e.g. is $\sigma$ tight?}
	
Let us continue Example~\ref{ex:notAsolvable}, where $\mathbb D$ is $\mathbb Z$,
$\mathcal C$ is LIA, and $\mathcal A$ is 2VAR. 
Let $P_0$ be the set consisting of clauses 2--6. 

Starting from $P_{0}$ we construct 
a transformation sequence $P_{0}, P_{1}, P_{2}, P_{3}, P_{4}$, as we now indicate.
First, by applying the definition rule, we introduce the following new  predicate:

\vspace{1mm}
\noindent
\makebox[6mm][l]{7.}$su\_sq(M,R0,Sum,N,S0,Sqr) \leftarrow M\!=\!Y,\ su(M,R0,Sum),\ sq(N,Y,S0,Sqr)$

\smallskip

\noindent
We derive the clause set $P_1=P_0\cup\{7\}$ and $Defs_1=\{7\}$.
Now, by unfolding the atoms $su$ and $sq$ of clause~7, and then performing some more unfoldings
of the derived atoms, we get:

%\comment{REVIEWER: p. 11 in the transfomrations: in clauses 10,11,13 should R1 = R0 +1 not be R1  = R0 + M? }

\vspace{1mm}
%unf.
\noindent
\makebox[6mm][l]{8.}$su\_sq(M,R0,Sum,N,S0,Sqr) \leftarrow M\!=\!Y,\ M\!\leq\!0,\ Sum\!=\!R0,\ Sqr\!=\!S0$

\noindent
\makebox[6mm][l]{9.}$su\_sq(M,R0,Sum,N,S0,Sqr) \leftarrow M\!=\!Y,\ M\!\leq\!0,\ Sum\!=\!R0,\ Y\!>\!0,\ $

\hspace*{18mm}$Y1\!=\!Y\!-\!1,\ S1\!=\!S0\!+\!N,\ sq(N,Y1,S1,Sqr)$
%UNSAT BODY

\noindent
\makebox[6mm][l]{10.}$su\_sq(M,R0,Sum,N,S0,Sqr) \leftarrow M\!=\!Y,\ M\!>\!0,\ M1\!=\!M\!-\!1,\ R1\!=\!R0\!+\!M,\ $

\hspace*{18mm}$Y\!\leq\!0,\ Sqr\!=\!S0,\ su(M1,R1,Sum)$
%UNSAT BODY

\noindent
\makebox[6mm][l]{11.}$su\_sq(M,R0,Sum,N,S0,Sqr) \leftarrow M\!=\!Y,\ M\!>\!0,\ M1\!=\!M\!-\!1,\ R1\!=\!R0\!+\!M,\ $\nopagebreak

\hspace*{18mm}$Y\!>\!0,\ Y1\!=\!Y\!-\!1,\ S1\!=\!S0\!+\!N,\ su(M1,R1,Sum),\ sq(N,Y1,S1,Sqr)$

\smallskip

\noindent
We get $P_2=P_0\cup\{8,9,10,11\}$ and $Defs_2=\{7\}$.
(Here and in the rest of the example, for reasons of conciseness, we feel free to 
avoid to list some intermediate CHC sets in the transformation sequence.)
By the constraint replacement rule~R4 we can remove clauses~9 and~10, whose bodies have unsatisfiable constraints,
and replace clauses~8 and~11~by:

\vspace{1mm}
\noindent
\makebox[6mm][l]{12.}$su\_sq(M,R0,Sum,N,S0,Sqr) \leftarrow M\!\leq\!0,\ Sum\!=\!R0,\ Sqr\!=\!S0$

\noindent
\makebox[6mm][l]{13.}$su\_sq(M,R0,Sum,N,S0,Sqr) \leftarrow M\!>\!0,\ M1\!=\!M\!-\!1,\ R1\!=\!R0\!+\!M,\ $

\hspace*{18mm}$S1\!=\!S0\!+\!N,\ M1\!=\!Y1,\ su(M1,R1,Sum),\ sq(N,Y1,S1,Sqr)$

\smallskip

\noindent
We get $P_3=P_0\cup\{12,13\}$ and $Defs_3=\{7\}$. 
Then, by the folding rule~R3,
we fold clause~4 (in $P_0$) using clause~7 and
we derive the following clause:

%\comment{REVIEWER: when clauses 14 and 15 are generated, please state clearly that 
%	clause 14 comes from 4 \& 7 and 15 form 13 \& 7. }
	
\smallskip

\noindent
\makebox[6mm][l]{14.}$false \leftarrow Sum\!>\!Sqr, M\!\geq\!0, M\!=\!N, R0\!=\!0, S0\!=\!0, su\_sq(M,R0,Sum,N,S0,Sqr)$

\smallskip

Finally, we fold clause~13 using clause~7 and
we derive the following clause:

\smallskip

\noindent
\makebox[6mm][l]{15.}$su\_sq(M,R0,Sum,N,S0,Sqr)\! \leftarrow\! M\!>\!0,\ M1\!=\!M\!-\!1,\ R1\!=\!R0\!+\!M,\ $

\hspace*{18mm}$S1\!=\!S0\!+\!N,\ su\_sq(M1,R1,Sum,N,S1,Sqr)$

\smallskip

\noindent
We get the final set of clauses $P_4=(P_0 \setminus \{4\}) \cup\{12,14,15\}$.
Now, it is easy to check that the symbolic interpretation that maps the 
atom $su\_sq(M,R0,Sum,N,S0,Sqr)$ to 
the 2VAR constraint $Sum\!\leq\!Sqr \vee R0\!>\!S0 \vee M\!>\!N$,
and the $su$ and $sq$ atoms to $true$, is a 2VAR-definable model of $P_4$.
{This check can be done by replacing the atoms in~$P_4$ by the
corresponding symbolic interpretations, and then verifying the validity
of the formulas obtained in that way by using an SMT solver for
linear arithmetic, such as the popular Z3 solver~\cite{DeB08}.}

%\comment{REVIEWER: p.11 could you comment how the check is done : by hand? by Z3? }

Let us now make  some remarks on the derivation above. 

(1)~The applications of
the transformation rules satisfy the hypothesis of Theorem~\ref{thm:D-sat}
and hence  $P_0$ is $\mathbb Z$-satisfiable if and only if $P_4$  is $\mathbb Z$-satisfiable
(indeed, they are both  $\mathbb Z$-satisfiable).

(2)~The fact that  $P_4$ has a 2VAR-definable model, while $P_0$ has no such model,
is due to the fact that the applications of the folding rule R3 are not reversible foldings, and
hence the transformation sequence $P_0,\ldots,P_4$
does not satisfy the hypothesis of Theorem~\ref{thm:C-compl}.
Indeed, clause 7 occurs in $Defs_3$, but not in $P_3$. 
More in general, the derivation of a CHC set with an $\mathcal A$-definable model
from a CHC set without an $\mathcal A$-definable model
is due to the introduction of new predicates and also to the derivation
(via non-reversible foldings) of clauses that
constitute recursive definitions of these new predicates. 

(3)~Finally, note that at every step during the transformation 
from $P_0$ to $P_4$ we have
handled linear constraints only. However, the introduction of the
new predicate $su\_sq(M,R0,Sum,N,S0,Sqr)$, defined in terms of the conjunction
of $su(M,R0,Sum)$ and $sq(N,Y,S0,Sqr)$, allows us to discover linear relations 
between $Sum$ and $Sqr$ without having to deal with nonlinear constraints.
\hfill$\Box$
\end{example}

\section{Predicate Pairing}
\label{sec:strategy}

In Section~\ref{sec:rules} we have seen that by the applying unfold/fold
rules, in some cases one may transform a given $\mathbb D$-satisfiable set of clauses 
which does {\it not} admit an \mbox{$\mathcal A$-definable} model,
into an equisatisfiable set of clauses which admits an \mbox{$\mathcal A$-definable} model.
Then, the \mbox{$\mathbb D$-satisfiability} of the set of clauses
can be proved by a CHC solver that 
constructs an $\mathcal A$-definable model.
Example~\ref{ex:notAsolvableCont} of Section~\ref{sec:rules} suggests that
the crucial step in that transformation is a predicate pairing step, that is,
the introduction of a new predicate, say $t$,
whose defining clause has in its body the conjunction
of two atoms, one with predicate, say $q$, and the other with predicate, say $r$,
whose definitions are provided by the original set of clauses.
Indeed, the predicate pairing allows us to derive suitable relations between arguments
of the conjunction of $q$ and $r$, which cannot be expressed by
using constraints on the arguments of~$q$ and~$r$ separately.

In general, in order to do the transformation and perform some required
folding steps, it may be necessary to introduce, by predicate pairing, more than one definition.
The introduction of these new definitions
is a major issue in the case where the predicates to be paired 
should be chosen 
from the various predicates occurring in the conjunction of several atoms %% alberto
%%from a conjunction of several predicates occurring
in the body of a clause.
In particular this issue arises when predicates are defined by
{\it nonlinear} clauses, and hence their {repeated} unfolding may generate unbounded conjunctions
of atoms.

In this section we will present a strategy, called 
{\it Predicate Pairing}, for making the choice of the predicates to be paired. 
This strategy, which 
is realized by Algorithm~\ref{alg:pairing1},
takes as input a 
set of clauses and derives a new, equisatisfiable set of clauses, 
which by Theorem~\ref{thm:A-sound} is guaranteed to admit an $\mathcal A$-definable model,
whenever the original clauses had one. 
%\comment{DELETE?: (As usual, in Algorithm~\ref{alg:pairing1} we assume that comma is associative and commutative.)}
Actually, as we will show,
in many interesting cases the Predicate Pairing strategy constructs
new sets of clauses for which a CHC solver is able to construct one such model,  
while the same solver is unable to do so for the original set of clauses.
We present the Predicate Pairing strategy with the help of an example.
Suppose that we are given the following specification of (a variant of) the Ackermann function:

%%%%%%%%%%%%%%%%%%%%%%%%%%%%%%%%%%%%%%%%%%%%%%%%%%%

% ---------------------------------------------------------------------------------
% Transformation strategy 1: left-to-right pairing + entailed equalities
% ---------------------------------------------------------------------------------
\SetAlCapSkip{40ex}
\SetAlCapSkip{30mm}
%\NoCaptionOfAlgo
\begin{algorithm}[ht]
%%% INPUT
\KwInput{\rule{0mm}{3.5mm}\hangindent=0mm(i)~a clause $C_{\small{init}}$ of the form:
${\textit false} \leftarrow c_{\small{init}}, q(X), r(Y)$, %where $c$ is a constraint, 
and 

(ii) two disjoint sets $\!Q$ and~$\!R$ of clauses such that:
$q$ occurs in $\!Q$, $r$ occurs in $\!R$, and 
$Q$ and~$R$ have no predicates in common.
}
%%% OUTPUT
\KwOutput{\hangindent=0mm a set $\!\textit{TransfCls}$ of clauses such that there is no occurrence of an atom
with predicate in $Q$ and an atom with predicate in $R$ in the same clause body.}

\vspace{-2.5mm}
\rule{30mm}{.5pt}

\vspace{-.5mm}
\noindent\hangindent=0mm %%% NOTATION
\noindent\hangindent=0mm
{\bf Notation}: 

For all constraints $d$, for all atoms $A$ and $B$,\\ let {\it Eq\/}$(d,\!A,\!B)$ 
be $\{X\!\!=\!\!Y ~\mid~ X\!\!\in\!\textit{vars}(A), ~~Y\!\!\in\!\textit{vars}(B), ~~~\mathbb{D} 
\models \forall(d\!\rightarrow\! (X\!\!=\!\!Y))\}$.\\

\vspace*{-6pt}\rule{70mm}{0.2mm}

$\textit{InCls}:= \{C_{\small{init}}\}$; ~~~$\textit{Defs}:= \emptyset$; ~~~$\textit{TransfCls}:= Q\cup R$\;
%%% OUTER WHILE
	\While{\rm in \textit{InCls} there is a clause~$C$ of the form: 
${L} \leftarrow {c}, {A}_{Q}, {B}_{R}$\,, where:\\ \hspace{10.5mm}(i)~${A}_{Q}$~occurs in $Q$, and 
(ii)~${B}_{R}$ occurs in $R$~ } {
		\vspace{1mm}
% UNFOLDING ========================
	   \LeftComment{\textsc{Unfolding}}: %\hspace{3.2mm}
	    \hangindent=3.5mm Unfold once the atom {${A}_{Q}$} and once the atom {${B}_{R}$} in clause~$\!C\!$ using 
	    $Q\cup\! R$, thereby deriving the set \textit{UnfoldedCls}  of clauses\,;

% DEFINITION and FOLDING ================
		\smallskip
		 \LeftComment{\textsc{Definition}\,\&\,\textsc{Folding}:}
		$\textit{FoldedCls}:= \textit{UnfoldedCls}$\,;\\[.7mm]

%%%% INNER WHILE
		\While{\rm in $\textit{FoldedCls}$ there is clause $E$ of the form: 
		     $H\!\leftarrow\!d, A, B, G$, where: (i)\,$A$ and\,$B$\,are atoms whose predicates occur in 
		     $\!Q$\,and $\!R$,
		     respectively, and\\ 
		     (ii) $\!G$ is\,a\,conjunction\,of atoms,\,such that:\\ for all atoms $\!M\!$ and $\!N$\,in\,$(A,\! B,\! G)$, whose predicates occur in $Q$ and $R$, respectively, 
		we have that $\mid\!{\textit Eq\/}(d,\!A,\!B)\!\mid\ \geq\ \mid\!{\textit Eq\/}(d,\!M,\!N)\!\mid~~$\vspace{1mm}
	    }{
	    \eIf{{\rm \,in \textit{Defs} there is a clause $D'$ of the form:  
			$H'\leftarrow\!d',{A}', {B}'$  such that,\\ for some substitution $\vartheta$, 
			we can fold $(A, B)$ in clause~$E$ using $D'$,\\$\!\!$}} 
            %% THEN
  {\vspace*{-4.5mm}~~~$\textit{FoldedCls}:=(\textit{FoldedCls}\setminus\{E\})~\cup~\{H\!\leftarrow\!d,H'\vartheta,G\}$\;}
            %% ELSE
            {let $D$ be the clause ~$newp(Z)\!\leftarrow\!{e},A,B$,  where:
		    \vspace*{-.3mm}\begin{enumerate}[label={(\roman*)},nolistsep,itemsep=0pt,align=right,labelsep=3pt,leftmargin=6.5mm]
	        \item \textit{newp} is a predicate symbol not occurring elsewhere,\\
	        \item $Z=\textit{vars}(A) \cup \textit{vars}(B)$, and\\ 
	        \item \textit{e} is the conjunction of the equalities in {\it Eq\/}$(d,\!A,\!B)$;
	        \end{enumerate}

            \vspace{.5mm}
			$\textit{FoldedCls}:=(\textit{FoldedCls}\setminus \{E\})~\cup~\{H\!\leftarrow\!d, \textit{newp}(Z),G\}$\;
			$\textit{Defs}:=\textit{Defs}~\cup~\{D\}$; ~~$\textit{InCls}:=\textit{InCls}\cup\{D\}$\;
		}
		%---- end else
	}
%%%% -----------------
	$\textit{InCls}:=\textit{InCls}\setminus\{C\}$; ~~$\textit{TransfCls}:=\textit{TransfCls}\cup \textit{FoldedCls}$\;  
	
}
\caption{\vspace*{.8mm}The {\it Predicate Pairing\/} strategy.$^{^{^{~}}}$
%\rule{0mm}{3.5mm}
%\raisebox{-2mm}{\rule{0mm}{6mm}}
}\label{alg:pairing1}
\end{algorithm}
%%%%%%%%%%%%%%%%%%%%%%%%%%%%%%%%%%%%%%%%%%%%%%%%%%%

\renewcommand{\baselinestretch}{.9}

\noindent
\begin{minipage}[t]{1\textwidth}

\makebox[104mm][l]{S1.\ $\textit{ackermann}(m,\!n) = n\!+\!1$}$\textit{if}\ m\!\leq\!0$ 

\makebox[104mm][l]{S2.\ $\textit{ackermann}(m,\!n) = \textit{ackermann}(m\!-\!1,1)$}$\textit{if}\  m\!>\!0\wedge n\!=\!0$

\makebox[104mm][l]{S3.\ $\textit{ackermann}(m,\!n) = \textit{ackermann}(m\!-\!1,\textit{ackermann}(m,n\!-\!1))$}$\textit{if}\ m\!>\!0 \wedge  n\!>\!0$ 
\end{minipage}

\renewcommand{\baselinestretch}{1} 

\vspace{2mm}
\noindent
and the following two programs each of which implements that specification:

%\noindent
\begin{minipage}[t]{1\textwidth}
{\small{
\begin{verbatim}
int ackermann1(int m, int n) {
    if (m =< 0) { return n+1; } 
    else if (m > 0 && n = 0) { ackermann1(m-1,1); }
    else if (m > 0 && n > 0) { ackermann1(m-1,ackermann1(m,n-1)); }
  }
\end{verbatim}
}}
\end{minipage}

\smallskip
%\noindent
\begin{minipage}[t]{1\textwidth}
{\small{\begin{verbatim}
int ackermann2(int m, int n) {
  while (m > 0) {
    if (n == 0) { m = m-1;   n = 1; } 
    else { n = ackermann2(m,n-1);   m = m-1; }
  }
  return n+1;
}
\end{verbatim}
}}
\end{minipage}

\vspace{1.5mm}
\noindent
We want to prove the equivalence of these two implementations, in the sense that, 
for all non-negative integers ${\mathtt{m\!\geq\!0}}$ and ${\mathtt{n\!\geq\!0}}$,  
{\texttt{ackermann1(m,n)}} returns the same integer returned by {\texttt{ackermann2(m,n)}}.

Given the programs {\texttt{ackermann1(m,n)}} and {\texttt{ackermann2(m,n)}},
we first generate the following two sets $Q$ and $R$ of clauses that
encode the operational semantics of the programs. These sets of clauses can be 
derived by specializing the interpreter of the imperative language with respect to the 
programs~\cite{De&17b}.

%%The first step of our proof consists in generating a set of constrained Horn clauses 
%%for the two programs.
%%This is done by using a familiar technique, 
%%called the
%%{\it verification condition generation}
%%(we do not enter into the details of this technique here, 
%%and we refer to~\cite{De&15b}). By doing so, 
%%we get the following two sets $Q$ and $R$ of clauses.

\newpage
\smallskip
\noindent
$Q\!:$~~\vspace*{-4.5mm}
\begin{enumerate}[topsep=0pt,parsep=1pt,partopsep=0pt,
		itemsep=-1pt,labelsep=4pt,leftmargin=30pt]
    \item\label{ack1} $\textit{ackermann\/}1(M1,N1,A1) \leftarrow ack1(M1,N1,A1)$\nopagebreak
	\item\label{ack2} $ack1(M1,N1,A1) \leftarrow  M1\!\leq\!0, A1\!=\!N1\!+\!1$
	\item\label{ack3} $ack1(M1,N1,A1) \leftarrow  M1\!>\!0, N1\!=\!0, X1\!=\!M1\!-\!1, Y1\!=\!1, 
	~ack1(X1,Y1,A1)$\nopagebreak
	\item\label{ack4} $ack1(M1,N1,A1) \leftarrow  M1\!>\!0, N1\!>\!0, X1\!=\!M1\!-\!1, Y1\!=\!N1\!-\!1,$ 
	
	~~~~$ack1(M1,Y1,Z1),\ ack1(X1,Z1,A1)$
\end{enumerate}

\smallskip
\noindent
%\begin{minipage}{1\textwidth}
$R\!:$~~\vspace*{-4.5mm}
\begin{enumerate}[topsep=0pt,parsep=1pt,partopsep=0pt,
		itemsep=-1pt,labelsep=4pt,leftmargin=30pt,resume]
	\item\label{ack5} $\textit{ackermann\/}2(M2,N2,A2) \leftarrow A3\!+\!1=A2, ack2(M2,N2,A3)$
	\item\label{ack6} $ack2(M2,N2,A2) \leftarrow  M2\!\leq\!0, A2\!=\!N2$
	\item\label{ack7} $ack2(M2,N2,A2) \leftarrow  M2\!>\!0, N2\!=\!0, M2\!=\!X2\!+\!1, Y2\!=\!1, ~ack2(X2,Y2,A2)$
	\item\label{ack8} $ack2(M2,N2,A2) \leftarrow  M2\!>\!0, N2\!\neq\!0, X2\!=\!M2\!-\!1, Y2\!=\!N2\!-\!1, Z2\!=\!Z3\!-\!1,$\nopagebreak
	
	~~~~$ack2(M2,Y2,Z2),\ ack2(X2,Z3,A2)$
\end{enumerate}
%\end{minipage}

\smallskip
\noindent
The equivalence of the functions computed by the programs for {\texttt{ackermann\/1}} 
and {\texttt{ackermann\/2}} is expressed   
in terms of the predicates defined by clauses~1--8 as follows: for all 
integers $M1$, $M2$, $N1$, $N2$, we have that if 
$M1\!\geq\!0$, \mbox{$M1\!=\!M2$}, \mbox{$N1\!\geq\!0$}, \mbox{$N1\!=\!N2$}, 
and 
$\textit{ackermann\/}1(M1,N1,A1)$ and 
$\textit{ackermann\/}2(M2,N2,A2)$ both hold, then 
\mbox{$A1\!=\!A2$} holds.
Thus, given the clause:

\smallskip
\begin{enumerate}[topsep=0pt,parsep=1pt,partopsep=0pt,
	itemsep=-1pt,labelsep=6pt,leftmargin=14pt,resume]
\item\label{ackquery} $\textit{false} \leftarrow A1\!\neq\!A2,\ M1\!\geq\!0,\ M1\!=\!M2,\ N1\!\geq\!0,\ N1\!=\!N2, $ 
\item[]	$\clspace \textit{ackermann\/}1(M1,N1,A1),\ \textit{ackermann\/}2(M2,N2,A2)$		
\end{enumerate}

\smallskip

\noindent
the proof that $\texttt{ackermann\/}1$ and
 $\mathtt{ackermann\/}2$ are equivalent is reduced to the construction of a model
for clauses~\ref{ack1}--\ref{ackquery} (note that the constraint $A1\!\neq\!A2$ in clause~\ref{ackquery} 
states that the values returned by the two programs are different).

Now we have that no  CHC solver that constructs LIA-definable models,
can prove the satisfiability of clauses~\ref{ack1}--\ref{ackquery}.
{Indeed, in order to make that proof, the solver should discover
that the atoms $\textit{ackermann\/}1(M1,N1,A1)$ and 
$\textit{ackermann\/}2(M2,N2,A2)$ imply the two equalities
$A1\!=\!\textit{ackermann\/}(M1,N1)$ and $A2\!=\!\textit{ackermann\/}(M2,N2)$, 
respectively, 
where $\textit{ackermann\/}$ is 
the  function specified by equations~$S1$--$S3$, 
and  these equalities cannot
be expressed as linear integer constraints.}

Thus, in order to allow a CHC solver to construct
a LIA-definable model for clauses~\ref{ack1}--\ref{ackquery}, one should avoid reasoning on
the two predicates $\textit{ackermann\/}1$ and $\textit{ackermann\/}2$  in a separate way, and instead, one should reason on the {\textit{conjunction}} 
of those predicates. Indeed, in what follows
we will derive for that conjunction a new,  equisatisfiable set of clauses 
that has a LIA-definable model. 
In this new set of clauses we will discover suitable 
LIA constraints relating the arguments of the predicates $\textit{ackermann\/}1(M1,N1,A1)$ and 
$\textit{ackermann\/}2(M2,N2,A2)$. Using these constraints the CHC solver Z3 
can show the existence of a LIA-definable model for clauses~\ref{ack1}--\ref{ackquery},
thereby proving the desired equivalence between programs
$\texttt{ackermann\/}1$ and
 $\mathtt{ackermann\/}2$.

This new set of clauses % which has a LIA-definable model,
will be derived from clause~\ref{ackquery} by applying a sequence of  transformation 
rules according to the Predicate Pairing strategy,
as indicated in 
Algorithm~\ref{alg:pairing1}. 
The algorithm takes as input a set of clauses $\{C_{\mathit{init}}\} \cup Q \cup R$, that is,
$\{\ref{ackquery}\} \cup \{\ref{ack1},\ref{ack2},\ref{ack3},\ref{ack4}\} \cup 
\{\ref{ack5},\ref{ack6},\ref{ack7},\ref{ack8}\}$
in our case, and produces as output 
a new set of clauses by applying the \textrm{unfolding}, \textrm{definition}, and \textrm{folding}~rules.
During the application of that strategy 
we silently apply the constraint replacement rule to remove clauses which have
unsatisfiable constraints in their body.

%%%%%%%%%%%%%%%%%%%%%%%%%%%%%%%%%%%%%%%%%
% ------------------------------------------------------------------------------
% first iteration +++

\medskip
\noindent
$\bullet$ {\it{First iteration of the body of the while-loop of Predicate Pairing.}}

Since $InCls=\{\ref{ackquery}\}$, $\textit{ackermann\/}1$ occurs in
$Q$, and $\textit{ackermann\/}2$ occurs in $R$, we start off by unfolding 
$\textit{ackermann\/}1(M1,N1,A1)$ and $\textit{ackermann\/}2(M2,N2,A2)$
in clause~\ref{ackquery}. These unfoldings correspond to a symbolic 
evaluation step of each of the two atoms. We get:

\begin{enumerate}[topsep=0pt,parsep=1pt,partopsep=0pt,
	itemsep=-1pt,labelsep=6pt,leftmargin=18pt,resume]
	\item\label{u1}   $\textit{false} \leftarrow A1\!\neq\!A2,\ M1\!\geq\!0,\ M1\!=\!M2,\ N1\!\geq\!0,\ N1\!=\!N2,\  A2\!=\!A3\!+\!1,$
	
	$\clspace ack1(M1,N1,A1), ack2(M2,N2,A3)$
\end{enumerate}

\noindent
Then in order to fold clause~\ref{u1}, we introduce the following definition clause~\ref{d1} which
pairs together the atoms with predicate $ack1$ occurring in $Q$ and predicate $ack2$ 
occurring in $R$. In the body of this definition we have the 
equality constraints $M1\!=\!M2$ and $N1\!=\!N2$ between the arguments of $ack1$ and $ack2$.

\begin{enumerate}[topsep=0pt,parsep=1pt,partopsep=0pt,
	itemsep=-1pt,labelsep*=6pt,leftmargin=18pt,resume]
\item\label{d1} $new1(M1,N1,A1,M2,N2,A2) \leftarrow M1\!=\!M2, N1\!=\!N2,$\nopagebreak

$\clspace ack1(M1,N1,A1), ack2(M2,N2,A2)$
\end{enumerate}

\noindent
The
definition of $new1$ is then used for replacing, by folding, 
the conjunction of the atoms with predicates $ack1$ and $ack2$ in 
the body of clause~\ref{u1}. Thus, from clause~\ref{u1}, by folding,
we derive:

\begin{enumerate}[topsep=0pt,parsep=1pt,partopsep=0pt,
	itemsep=-1pt,labelsep=6pt,leftmargin=18pt,resume]
\item\label{f1}   $\textit{false} \leftarrow A1\!\neq\!A2,\ M1\!\geq\!0,\ M1\!=\!M2,\  
N1\!\geq\!0,\ N1\!=\!N2,\ A2\!=\!A3\!+\!1,$\nopagebreak
 
$\clspace new1(M1,N1,A1,M2,N2,A3)$
\end{enumerate}

% ------------------------------------------------------------------------------
% second iteration +++
\smallskip
\noindent
$\bullet$ {\it{Second iteration of the body of the while-loop of Predicate Pairing.}}

Since $InCls=\{\ref{d1}\}$, $\textit{ack\/}1$ occurs in
$Q$, and $\textit{ack\/}2$ occurs in $R$, we have to perform a second iteration 
of the body of the while-loop of the Predicate Pairing.

We unfold 
the atoms with predicate $ack1$ and $ack2$ in the premise of 
 clause~\ref{d1}, and we get the following three clauses:

%%{Now we perform a second iteration of the while-loop of Predicate Pairing strategy
%%%Algorithm \ref{alg:pairing1}
%%starting from the newly introduced clause~\ref{d1}. }
%%The atoms with predicate $ack1$ and $ack2$ in the premise of 
%% clause~\ref{d1} are unfolded, and the following three clauses are derived:

\begin{enumerate}[topsep=0pt,parsep=0pt,partopsep=1pt,
	itemsep=0pt,labelsep*=4pt,leftmargin=18pt,resume]
\item\label{u2} $new1(M1,N1,A1,M2,N2,N2) \leftarrow\!M1\!\leq\!0, 
M1\!=\!M2, N1\!=\!N2, A1\!=\!N1\!+\!1$ 

\item\label{u3} $new1(M1,N1,A1,M2,N2,A2) \leftarrow M1\!>\!0, 
M1\!=\!M2, N1\!=\!0, N1\!=\!N2, $ 

$\clspace\!X1\!=\!M1\!-\!1, Y1\!=\!1, X2\!=\!M2\!-\!1, Y2\!=\!1, 
ack1(X1,\!\!Y1,\!A1), ~ack2(X2,\!\!Y2,\!A2)$
 
\item\label{u4} $new1(M1,N1,A1,M2,N2,A2)\leftarrow M1\!>\!0, 
M1\!=\!M2, N1\!>\!0, N1\!=\!N2, N2\!\neq\!0, $

$\clspace\!Y1\!=\!N1\!-\!1, X1\!=\!M1\!-\!1, Y2\!=\!N2\!-\!1, X2\!=\!M2\!-\!1,
Z3\!=\!Z2\!+\!1,$

$\clspace\!ack1(M1,Y1,Z1), ~ack1(X1,Z1,A1), ~ack2(M2,Y2,Z2), ~ack2(X2,Z3,A2)$ 
\end{enumerate}

\noindent
Clause~\ref{u2} need not be folded because it has no atoms in its body. Clause~\ref{u3} 
can be folded using 
clause~\ref{d1} (the conditions for folding given in Section~\ref{sec:rules} are indeed satisfied),
and we get:

\begin{enumerate}[topsep=0pt,parsep=0pt,partopsep=1pt,
	itemsep=0pt,labelsep*=4pt,leftmargin=18pt,resume]
\item\label{f3} $new1(M1,N1,A1,M2,N2,A2) \leftarrow M1\!=\!M2, M1\!>\!0, N1\!=\!N2, N1\!=\!0,$ 

$\clspace X1\!=\!M1\!-\!1, Y1\!=\!1, X2\!=\!M2\!-\!1, Y2\!=\!1, new1(X1,Y1,A1,X2,Y2,A2)$ 
\end{enumerate}

\noindent
Clause~\ref{u4} should be folded, but first we  need to choose the atoms
to be paired together.
According to our goal of proving a relation between the arguments of
$\textit{ackermann}1$ and $\textit{ackermann}2$, 
we should pair together an $ack1$ atom with an $ack2$ atom. 
However, the choice of the atoms to be paired can be made in different ways
because in clause~15 there are two $ack1$ atoms and two $ack2$ atoms.
The strategy we propose looks at the arguments of the atoms and selects the two
atoms which share a maximal number of equality constraints holding
between an argument of $ack1$ and an argument of $ack2$.
 
According to this strategy we have that $ack1(M1,Y1,Z1)$ should be paired
with $ack2(M2,Y2,Z2)$ because these two atoms share the two equalities
$M1\!=\!M2$ and $Y1\!=\!Y2$ (this last equality follows from
 $N1\!=\!N2$), while $ack1(M1,Y1,Z1)$ shares no equalities with 
$ack2(X2,Z3,A2)$. Moreover,
$ack1(X1,Z1,A1)$ shares one equality only, namely $X1\!=\!X2$ (this equality follows from
$M1\!=\!M2$) with $ack2(X2,Z3,A2$). 
Thus, we pair $ack1(M1,Y1,Z1)$ with
$ack2(M2,Y2,Z2)$ and then we take clause~\ref{d1}
for folding these two atoms. 

In order to fold the other two 
atoms occurring in clause~\ref{u4}, that is,
$ack1(X1,\!Z1,\!A1)$ 
and $ack2(X2,Z3,A2)$, 
we introduce the following clause:

\begin{enumerate}[topsep=0pt,parsep=0pt,partopsep=1pt,
	itemsep=-1pt,labelsep*=3pt,leftmargin=16pt,resume]
\item\label{d2}  $new2(M1,\!N1,\!A1,\!M2,\!N2,\!A2)\! \leftarrow\! M1\!=\!M2, ack1(M1,\!N1,\!A1), ack2(M2,\!N2,\!A2)$

%$\clspace  ack1(M1,N1,A1), ack2(M2,N2,A2)$ 
\end{enumerate}

\noindent
Thus, by folding clause~\ref{u4} using clauses~\ref{d1} and~\ref{d2}, we get:\nopagebreak

\begin{enumerate}[topsep=0pt,parsep=0pt,partopsep=1pt,
	itemsep=-1pt,labelsep*=4pt,leftmargin=18pt,resume]
\item\label{f4} $new1(M1,N1,A1,M2,N2,A2)\leftarrow M1\!=\!M2, M1\!>\!0, N1\!=\!N2, N1\!>\!0, N2\!\neq\!0, $ 
\item[] $\hspace{5pt}Y1\!=\!N1\!-\!1, X1\!=\!M1\!-\!1, Y2\!=\!N2\!-\!1, X2\!=\!M2\!-\!1, Z3\!=\!Z2\!+\!1,$
\item[] $\hspace{5pt}new1(M1,Y1,Z1,M2,Y2,Z2),\ 
new2(X1,Z1,A1,X2,Z3,A2)$ 
\end{enumerate}

\noindent
The basic idea of our pairing strategy %for introducing new definitions
is that the atoms that are paired together, having some of their arguments equal, 
have a somewhat synchronized behavior and 
this synchronization may determine, for the other arguments, the existence of simple relations 
that are easy to express in the theory of constraints one considers.

\smallskip
At this point of the application of the Predicate Pairing strategy
%algorithm 
we have that 
$\textit{TransfCls} =\!Q\cup R \cup \{\ref{f1},
\ref{u2},\ref{f3},
\ref{f4}\}$, $\textit{Defs}=\{\ref{d1},\ref{d2}\}$,
and $\textit{InCls}=\{\ref{d2}\}$.

%% third iteration ----------
\smallskip
\noindent
$\bullet$ {\it{Third iteration of the body of the while-loop of Predicate Pairing.}}
%\smallskip

Since $InCls\!=\!\{\ref{d2}\}$ and 
 clause~\ref{d2} has the atom $ack1$ that occurs in $Q$ and 
the atom $ack2$ that occurs in~$R$,
we have to perform a new iteration of the body of the while-loop of the Predicate Pairing.
%Algorithm~\ref{alg:pairing1}.

In clause~\ref{d2} we unfold once the atom with predicate $ack1$ and the atom with predicate $ack2$, 
and we get:

\begin{enumerate}[topsep=0pt,parsep=0pt,partopsep=1pt,
	itemsep=-1pt,labelsep*=4pt,leftmargin=18pt,resume]
%setcounter{enumi}{18}
% clause 19
\item\label{u5} $new2(M1,N1,A1,M2,N2,N2) \leftarrow M1\!=\!M2, M1\!\leq\!0, A1\!=\!N2\!+\!1$ 

% clause 20
\item\label{u6} $new2(M1,N1,A1,M2,N2,A2) \leftarrow M1\!=\!M2, M1\!>\!0, N1\!=\!0, N2\!=\!0,$

$\clspace X1\!=\!M1-1,\! Y1\!=\!1, X2\!=\!M2-1,\! Y2\!=\!1, ack1(X1,\!Y1,\!A1), ack2(X2,\!Y2,\!A2)$ 

% clause 21
\item\label{u7} $new2(M1,N1,A1,M2,N2,A2) \leftarrow M1\!=\!M2, M1\!>\!0, N1\!=\!0, N2\!\neq\!0,$\nopagebreak

$\clspace X1\!=\!M1\!-\!1, Y1\!=\!1, X2\!=\!M2\!-\!1, Y2\!=\!N2\!-\!1, 
Z3\!=\!Z2\!+\!1,$\nopagebreak

$\clspace ack1(X1,Y1,A1),\ ack2(M2,Y2,Z2),\ ack2(X2,Z3,A2)$ 

% clause 22
\item\label{u8} $new2(M1,N1,A1,M2,N2,A2) \leftarrow M1\!=\!M2, M1\!>\!0, N1\!>\!0, N2\!=\!0,$ 

$\clspace X1\!=\!M1\!-\!1, Y1\!=\!N1\!-\!1, X2\!=\!M2\!-\!1, Y2\!=\!1,$

$\clspace ack1(M1,Y1,Z1), ack1(X1,Z1,A1), ack2(X2,Y2,A2)$ 

% clause 23
\item\label{u9} $new2(M1,N2,A1,M2,N2,A2) \leftarrow M1\!=\!M2, M1\!>\!0, N1\!>\!0, N2\!\neq\!0,$

$\clspace X1\!=\!M1\!-\!1, Y1\!=\!N2\!-\!1, X2\!=\!M2\!-\!1, Y2\!=\!N2\!-\!1, Z2+1=Z3,$ 

$\clspace ack1(M1,Y1,Z1), ack1(X1,Z1,A1), ack2(M2,Y2,Z2), ack2(X2,Z3,A2)$
\end{enumerate}

\noindent
Clause~\ref{u5} need not be folded. Clause~\ref{u6} can be folded using definition~\ref{d2} and we get:

\begin{enumerate}[topsep=0pt,parsep=0pt,partopsep=2pt,
	itemsep=0pt,labelsep*=4pt,leftmargin=18pt,resume]
\setcounter{enumi}{23}
\item\label{f6}  $new2(M1,N1,A1,M2,N2,A2) \leftarrow M1\!=\!M2, M1\!>\!0, N1\!=\!0, N2\!=\!0,$ 

$\clspace X1\!=\!M1\!-\!1, Y1\!=\!1,  X2\!=\!M2\!-\!1, Y2\!=\!1, new2(X1,Y1,A1,X2,Y2,A2)$ 
\end{enumerate}

\noindent
In order to fold clause~\ref{u7}, first we select the two atoms with the predicates to be paired. 
%% alberto
We have that $ack1(X1,Y1,A1)$ shares one equality with $ack2(X2,Z3,A2)$, that is,
$X1\!=\!X2$ (this equality follows from $M1\!=\!M2$),    and shares no equalities with $ack2(M2,\!Y2,\!Z2)$. Hence we select the atoms  $ack1(X1,\!Y1,A1)$ and $ack2(X2,Z3,A2)$ 
in the body of clause~\ref{u7}, and we fold that clause by using clause~\ref{d2}, thereby deriving the following clause:

\begin{enumerate}[topsep=0pt,parsep=0pt,partopsep=2pt,
	itemsep=0pt,labelsep*=4pt,leftmargin=18pt,resume]
	%\setcounter{enumi}{10}
% clause 25
\item\label{f7}  $new2(M1,N1,A1,M2,N2,A2) \leftarrow M1\!=\!M2, M1\!>\!0, N1\!=\!0, N2\neq0,$\nopagebreak

$\clspace X1\!=\!M1\!-\!1, Y1=1,X2\!=\!M2\!-\!1, Y2\!=\!N2\!-\!1, Z3\!=\!Z2\!+\!1,$\nopagebreak

$\clspace  new2(X1,Y1,A1,X2,Z3,A2), ack2(M2,Y2,Z2)$ 
\end{enumerate}

\noindent 
%By processing clauses~\ref{u8} and~\ref{u9} in a similar manner, we get:

\noindent
%\begin{minipage}[t]{1\textwidth}
\begin{enumerate}[topsep=0pt,parsep=0pt,partopsep=2pt,
	itemsep=0pt,labelsep*=4pt,leftmargin=18pt,resume]

\item[] \hspace*{-6mm}By processing clauses~\ref{u8} and~\ref{u9} in a similar manner, we get:
\vspace{.5mm}

% clause 26
\item\label{f8}  $new2(M1,N1,A1,M2,N2,A2) \leftarrow M1\!=\!M2, M1\!>\!0, N1\!>\!0, N2\!=\!0,$\nopagebreak 

$\clspace  X1\!=\!M1\!-\!1, Y1\!=\!N1\!-\!1, X2\!=\!M2\!-\!1, Y2\!=\!1,$\nopagebreak 

$\clspace new2(X1,Z1,A1,X2,Y2,A2), ack1(M1,Y1,Z1)$

% clause 27
\item\label{f9} $new2(M1,N1,A1,M2,N2,A2) \leftarrow M1\!=\!M2, M1\!>\!0, N1\!>\!0, N2\!\neq\!0,$\nopagebreak 

$\clspace  X1\!=\!M1\!-\!1, Y1\!=\!N1\!-\!1, X2\!=\!M2\!-\!1, Y2\!=\!N2\!-\!1, Z3\!=\!Z2\!+\!1,$\nopagebreak 

$\clspace  new2(M1,Y1,Z1,M2,Y2,Z2), new2(X1,Z1,A1,X2,Z3,A2)$
\end{enumerate}
%\end{minipage}
\vspace{1mm}

\noindent
Since $\textit{InCls}\!=\!\emptyset$ no new iteration of the body
of the while-loop of the Predicate Pairing is required. Thus, the
application of that strategy terminates. 
The resulting set of clauses is
$\textit{TransfCls} =\!Q\cup R \cup 
\{\ref{f1},\ref{u2},\ref{f3},\ref{f4},
\ref{u5},\ref{f6},\ref{f7},\ref{f8},\ref{f9}\}$.

Now, the CHC solver Z3, when given as input the set $\textit{TransfCls}$ of 
clauses, constructs a \mbox{LIA-definable} model of $\textit{TransfCls}$. In particular, it
constructs a \mbox{LIA-defin}\-able model of clause~\ref{f1} by inferring that
 $new1(M1,\!N1,\!A1,\!M2,\!N2,\!A3)$ implies
$A1\!=\!A3\!+\!1$, which together with $A2\!=\!A3\!+\!1$, implies $A1\! =\!A2$,
and hence the body of the clause is shown to be false.

By Theorem~\ref{thm:D-sat}
the existence of a LIA-definable model for $\textit{TransfCls}$ entails that 
clauses~~\ref{ack1}--\ref{ackquery} have a $\mathbb Z$-model, and this concludes the
proof that programs \texttt{ackermann}1 and \texttt{ackermann}2 are equivalent.

\medskip
We end this section by stating
some results about the Predicate Pairing strategy.
First, we have that Predicate Pairing always terminates because the number
of new predicate definitions that can be introduced
is bounded by the number~$k$ of different
conjunctions of the form 
$(e,A,B)$,  where $A$ and $B$ are atoms whose predicates 
occur in $Q\cup R$, and ${e}$ is a conjunction of 
equalities between variables in $(A,B)$.
Hence, the number of executions of the body
of the while-loop of Predicate Pairing is at most $k$.

{It is easy to see that the sequence of applications of the transformation 
rules performed by the Predicate Pairing strategy
%Algorithm~\ref{alg:pairing1} 
constructs a transformation sequence where every definition in $\textit{Defs}$ is unfolded at least once.}
Thus, from Theorem~\ref{thm:D-sat}, which states
the equivalence with respect to $\mathbb D$-satisfiability, and 
Theorem~\ref{thm:A-sound}, which states the preservation of 
$\mathcal A$-definable models, we get the following result.

\begin{theorem}[Termination and soundness of the Predicate Pairing strategy]
\label{thm:soundness} 
Let the sets $\{C_{\mathit{init}}\}$, $Q$, and $R$ of clauses be the input of the 
Predicate Pairing 
strategy. Then the strategy terminates and returns a set 
{\it TransfCls} of clauses 
such that:

\noindent
(i) $\{C_{\mathit{init}}\} \cup Q \cup R$ is $\mathbb D$-satisfiable 
iff {\it TransfCls} is $\mathbb D$-satisfiable, and

\noindent
(ii) if $\{C_{\mathit{init}}\} \cup Q \cup R$ has an $\mathcal A$-definable model, 
then {\it TransfCls} has an $\mathcal A$-definable model.
\end{theorem}

Finally, note that the application of the Predicate Paring strategy may be iterated,
and hence, at the end of the transformation of a set of clauses, 
 more than two predicates may turn out to be tupled together.

%%%%%%%%%%%%%%%%%%%%%%%%%%%%%%%%%%%%%%%%%%%

\section{Case Studies: Relational Program Properties}
\label{sec:case_studies}
%\comment{As already mentioned, Predicate Pairing 
%can be particularly useful when we want to
%prove relational properties,
%that is properties that relate two executions of the same program, or of different programs.
%}
%%
In this section we illustrate the application of the Predicate
Pairing strategy to some relevant 
classes of relational program properties. In particular,
we have considered the following classes of properties: 
(i)~the equivalence of programs implementing nonlinear {and/or nested} recursive functions, 
(ii.1)~the injectivity of programs, (ii.2)~the monotonicity of programs,
(ii.3)~the functional dependence of programs,
(iii)~non-interference of programs, 
% an information-flow security property,
(iv)~equivalence of loop-optimized versions of programs with respect to the 
corresponding non-optimized versions, and
(v)~the equivalence of programs that manipulate integer arrays.
We will consider these classes of properties in separate subsections.

%\smallskip
%\comment{
%Now we briefly show how to encode relational properties  
%between predicates $p(A,B)$ and $q(X,\!Y)$   encoding 
% executions of programs {\it P} and {\it Q},  
%where $A$ and $X$ represent input values,
%and output values are represented by $B$ and $Y$, respectively.
%%
%Let us consider the relational property 
%stating that, if the constraint $\textit{pre}(A,\!B)$ holds before the execution of {\it P} and {\it Q},
%then the constraint $\textit{post}(A,B,X,\!Y)$ holds after their execution.
%%
%This property can be checked by testing the satisfiability of  
%the clauses defining predicates $p$ and $q$
%together with the following clause
%
%$\textit{false}  \leftarrow   \textit{pre}(A,B),\  \textit{notpost}(A,B,X,\!Y),  \ p(A,B),\ q(X,\!Y) $
%
%\noindent
%where $\textit{notpost}(A,B,X,\!Y)$ is a constraint
%which is equivalent to the negation of $\textit{post}(A,B,X,\!Y)$.\footnote{
%If the constraint domain is not closed under negation then 
%multiple satisfiability tests might be required
%for checking the relational property.} 
%
%A common relational property, \textit{program} (or \textit{predicate}) \textit{equivalence},
%can be obtained by instantiating the constraints 
%$\textit{pre}(A,B)$,
%$\textit{post}(A,B,X,\!Y)$, and
%$\textit{notpost}(A,B,X,\!Y)$
%with
%$A\!=\!B$, $X\!=\!Y$, and $X\!\neq\!Y$, respectively.
%
%}
%

	Now we briefly show how to encode relational properties  
	between the executions of two programs {\it P} and {\it Q}~\cite{De&16c}.
	We assume that the operational semantics of programs {\it P} and {\it Q}
	is represented by predicates $p(A,B)$ and $q(X,\!Y)$,  
	where $A$ and $X$ represent (tuples of) input values,
	and $B$ and $Y$ represent (tuples of) output values, respectively.
	As already mentioned in the previous sections, 
	the clauses defining $p$ and $q$ can be 
	derived by specializing the interpreter of the imperative language with respect to the 
	programs~\cite{De&17b}.

	Let us now consider the relational property 
	stating that, if the constraint $\textit{pre}(A,\!B)$ holds before the execution 
	of {\it P} and {\it Q} and the execution of these programs terminates,
	then the constraint $\textit{post}(A,B,X,\!Y)$ holds after the execution.
	This property can be verified by testing the satisfiability of  
	the CHC set consisting of the clauses defining predicates $p$ and $q$,
	together with the following clause:
	
\smallskip
{\it RP\/}: \ $\textit{false}  \leftarrow   \textit{pre}(A,B),\  \textit{notpost}(A,B,X,\!Y),  \ p(A,B),\ q(X,\!Y) $
	
\smallskip
	
\noindent
where $\textit{notpost}(A,B,X,\!Y)$ is a constraint
which is equivalent to the negation of $\textit{post}(A,B,X,\!Y)$\footnote{
If the constraint language has no negation symbol, but the negation of
a constraint is equivalent to a disjunction of constraints, {as in the case of
LIA,} then the relational property can be encoded by a set of clauses.}.
	
%	A common relational property is {{\it program  equivalence}, which
%	is encoded by instantiating the constraints 
%	$\textit{pre}(A,B)$ and
%%	$\textit{post}(A,B,X,\!Y)$, and
%	$\textit{notpost}(A,B,X,\!Y)$
%	to
%	\comment{add c(A) ?}$A\!=\!X$
%	%$B\!=\!Y$, 
%	and $B\!\neq\!Y$, respectively.
	
%\smallskip
The application of our method based on the use of the Predicate Pairing 
strategy, is often crucial for solving satisfiability problems that encode relational program properties.
In Section~\ref{sec:experiments} we will discuss the results we have obtained in an extensive experimental evaluation that we have conducted. %on several problems. 

\subsection{Functions with Nonlinear and/or Nested Recursion}\label{subsec:nlin}
Similarly to what has been considered in Section~\ref{sec:strategy},
where we have presented two imperative programs implementing 
the Ackermann function specification and then proved 
their equivalence, in this section we consider various equivalence problems 
for pairs of imperative programs implementing
some functional specifications. 
In each pair one imperative program uses recursion only and the
other one uses recursion and iteration.

The operational semantics of the two imperative programs is encoded using 
two distinct sets of CHCs, each defining a predicate for each program.
The recursive structure of these predicate definitions  
mirrors the control flow of two imperative programs.

\smallskip
%Let $f(x)$ 
%be a generic recursive function specification, and 
%$\mathit{p}1$ and $\mathit{p}2$ be the two predicates encoding 
%the operational semantics of two imperative program which
%implement the specification~$f(x)$\footnote{
%For reasons of simplicity here we consider a function specification 
%with a single argument only. The generalization to functions with more than one argument 
%is straightforward and we leave it to the reader.}.
%%%a generic recursive function $f(x)$\footnote{
%%%For reasons of simplicity we consider a function with a single argument. The
%%%generalization to functions with more arguments is straightforward.}.
%The equivalence property, stating that $\mathit{p}1$ and $\mathit{p}2$ 
%define the same input/output relation,
%holds if the following clause, together with the set of clauses 
%defining the predicates $\mathit{p}1$ and $\mathit{p}2$, is satisfiable:
%
%$\textit{false} \leftarrow X1\!=\!X2,\  Y1\!\neq\!Y2,\ \mathit{p}1(X1,\!Y1),\ \mathit{p}2(X2,\!Y2)$ 
%
%\noindent
%where $X1$, $X2$ correspond to input values
%and $Y1$, $Y2$ correspond to output values. 

Let us consider two predicates
$\mathit{p}1$ and $\mathit{p}2$ that encode
the operational semantics of the two imperative programs, say $P1$ and $P2$,
implementing a given function specification.
%%a generic recursive function $f(x)$\footnote{
%%For reasons of simplicity we consider a function with a single argument. The
%%generalization to functions with more arguments is straightforward.}.
The {{\it equivalence} between $P1$ and $P2$ holds if, 
under some precondition on the input values,
$\mathit{p}1$ and $\mathit{p}2$ 
define the same input/output relation.
This property holds if the following clause, together with the set of clauses 
defining the predicates $\mathit{p}1$ and $\mathit{p}2$, is satisfiable:

\smallskip

{\it EQ\/}: \ $\textit{false} \leftarrow c(X1),\ X1\!=\!X2,\  Y1\!\neq\!Y2,\ \mathit{p}1(X1,\!Y1),\ \mathit{p}2(X2,\!Y2)$ 

\smallskip

\noindent
where: (i)~$X1$, $X2$ represent tuples of input values,
(ii)~$Y1$, $Y2$ represent tuples of output values, and 
(iii)~$c(X1)$ is a precondition on the input values. 
The reader may note that clause {\it EQ} is an instance of clause {\it RP} defining the general
relational property.
Note also that clause~\ref{ackquery} in Section~\ref{sec:strategy}, encoding the equivalence relation between the two implementations of the Ackermann function, is an instance of {\it EQ}.

%A common relational property is {\em program  equivalence}, which
%is encoded by instantiating the constraints 
%$\textit{pre}(A,B)$ and
%%	$\textit{post}(A,B,X,\!Y)$, and
%$\textit{notpost}(A,B,X,\!Y)$
%to
%\comment{add c(A) ?}$A\!=\!X$
%%$B\!=\!Y$, 
%and $B\!\neq\!Y$, respectively.
%

\smallskip

We have considered equivalence problems for imperative programs implementing
\textit{nonlinear} recursive functional specifications, 
that is, functional specifications with two or more recursive calls 
that depend on the same call (as in the case of the Fibonacci function).
Also, several of these specifications have \textit{nested recursions}, that is, 
they have recursive calls that are arguments of other recursive calls 
(as in the case of the Ackermann function),
thus making the verification problem more challenging.

In particular, in our experiments we have considered the following 
specifications of variants of
the Ackermann function\footnote{\tt http://mrob.com/pub/math/ln-2deep.html}. 
(Here and in the other function definitions we assume that $x,y,$ and $z$ are non-negative integers.)

\vspace{1mm}
\noindent
(1)~Original version by W.~Ackermann:

~~~$A(0,y,z)= y\!+\!z$, ~~~~$A(1,y,0)=0$, ~~~~$A(2,y,0)=1$,

~~~$A(x\!+\!3,y,0) = y$, 
~~~~$A(x\!+\!1,y,z\!+\!1) = A(x,y,A(x\!+\!1,y,z))$
 
\vspace{1mm}
\noindent
(2)~Variant by H.~Edelsbrunner:

~~~$E(0,y,z)= y\!+\!z$, ~~~~$E(x\!+\!1,y,0)=0$, ~~~~$E(x\!+\!1,y,1)=y$, 

~~~$E(x\!+\!1,y,z\!+\!2) = E(x,y,E(x\!+\!1,y,z\!+\!1))$

%%(i)~$A(0,y,z)= y\!+\!z$, (ii)~$A(x\!+\!1,y,0)=0$, (iii)~$A(x\!+\!1,y,1)=y$, \\ 
%%(iv)~$A(x\!+\!1,y,z\!+\!2) = A(x,y,A(x\!+\!1,y,z\!+\!1))$

%%$A(x,y,z)$ is 
%%(i) $y\!+\!z$ if $x=0$, or else
%%(ii) $0$  if $z=0$ % $x\!>\!0$ and $z=0$, 
%%(iii) $y$  if $z=1$ % $x\!>\!0$ and $z=1$, 
%%(iv) $A(x\!-\!1,y, A(x,y, z\!-\!1))$  otherwise.  
\vspace{1mm}

\noindent
(3)~Variant by R.~Robinson (we have used this variant in Section~\ref{sec:strategy}):
%which is probably the most popular version, 

~~~$R(0,y)= y\!+\!1$, ~~~~$R(x\!+\!1,0)= R(x,1)$,  ~~~~$R(x\!+\!1,y\!+\!1) = R(x,R(x\!+\!1,y))$

%%(i)~$A(0,y)\!=\! y\!+\!1$, (ii)~$A(x+1,0)\!=\! A(x,1)$,  (iii)~$A(x\!+\!1,y\!+\!1) \!=\! A(x,A(x\!+\!1,y))$

%%(i) $y\!+\!1$ if $x\!=\!0$, or else
%%(ii) $A(x\!-\!1,1)$ if $y\!=\!0$ % $x\!>\!0$ and $y\!>\!0$, 
%%(iii) $A(x\!-\!1,A(x,y\!-\!1))$ otherwise. % if $x\!>\!0$ and $y\!=\!0$. 
%%%$A(m,n)\!=\!$ if ($m\!=\!0$) \{$n\!+\!1$\} else if ($n\!>\!0$) \{$A(m\!-\!1,1)$\} else \{$A(m\!-\!1,A(m,n\!-\!1))$\} 

\vspace{1mm}
\noindent
(4)~Variant by R.~P\'eter: %R\'osza P\'eter, 

~~~$P(0,y)= 2y\!+\!1$, ~~~~$P(x+1,0) = P(x,1)$,  ~~~~$P(x\!+\!1,y\!+\!1) = P(x,P(x\!+\!1,y))$

%%(i)~$A(0,y)= 2y\!+\!1$, (ii)~$A(x+1,0) = A(x,1)$,  (iii)~$A(x\!+\!1,y\!+\!1) = A(x,A(x\!+\!1,y))$

%%$A(x,y)$ is 
%%(i) $2y\!+\!1$ if $x=0$, or else
%%(ii) $A(x\!-\!1,1)$ if $y\!>\!0$  
%%(iii) $A(x\!-\!1,A(x,y\!-\!1))$ otherwise.  

\smallskip
\noindent
Note that, these variants of the specification of the Ackermann function actually 
correspond to pairwise different functions. 
Indeed, we have that $A(2,y,0)\!\neq\! E(2,y,0)$  for all $y\!\geq\!0$, %  $A(2,y,0)=1$ while $E(2,y,0)=0$,
and $R(0,y)\!\neq \!P(0,y)$ for all $y\!>\!0$.  %$R(0,y)= y\!+\!1$  $P(0,y)= 2y\!+\!1$ 

%As we mentioned above, for each function specification (1)--(4), we have successfully 
%proved the equivalence of two imperative programs which implement that specification.

\smallskip
Additionally, we have considered some other equivalence problems 
for pairs of imperative programs encoding  % for
the following functional specifications:

\noindent
(5) a variant of 
the Sudan function:

~~~$S(0,y,z)= y\!+\!z$, ~~~~$S(x\!+\!1,y,0)=S(x,y\!+\!1,0)$,

~~~$S(x\!+\!1,y,z\!+\!1)= S(x,S(x\!+\!1,y,z),y\!+\!S(x\!+\!1,y,z))$

%%
%%(i)~$S(0,y,z)= y\!+\!z$, (ii)~$S(x\!+\!1,y,0)=S(x,y\!+\!1,0)$,\\ 
%%(iii)~$S(x\!+\!1,y,z\!+\!1)= S(x,S(x\!+\!1,y,z),y\!+\!S(x\!+\!1,y,z))$

%%$S(x,y,z)$ is 
%%(i) $y\!+\!z$ if $x=0$, or else
%%%(ii) $x$  if $z=0$   %% SUDAN
%%(ii) $S(x\!-\!1,y\!+\!1,0)$  if $z=0$    % FF version
%%(iii) $S(x\!-\!1, S(x,y,z\!-\!1) ,  y\!+\!S(x,y,z\!-\!1))$  otherwise;

\vspace{1mm}
\noindent
(6) the $B$ function:

~~~$B(0,y)= y\!+\!1$, ~~~~$B(x\!+\!1,y) = B(x,B(x\!+\!1,y\!-\!1))$

%%$B(x,y)$ is 
%%(i) $y\!+\!1$ if $x=0$ 
%%(iii) $B(x\!-\!1,B(x,y\!-\!1))$ otherwise;
%%
%%$B$ should be defined for all x, y $\geq$  0. $B(1,0)$ non termina in call-by-value. 
\vspace{1mm}
\noindent
(7) the $G$ function:
%%\comment{$G(x,y)$ is 
%%(i) $y\!+\!3$ if $x=1$, or else
%%(ii) $G(x\!-\!1,1)$ if $y\!>\!0$  
%%(iii) $G(x\!-\!1,G(x,y\!-\!1))$ otherwise; *** per $x$ = 0 ?? }
%%
%%

~~~$G(1,y)= y\!+\!3$, ~~~~$G(x\!+\!2,0)=G(x\!+\!1,1)$, 
~~~$G(x\!+\!2,y\!+\!1)=  G(x\!+\!1,G(x\!+\!2,y))$

%\vspace{1mm}
%\noindent
%as well as the following ones:\nopagebreak
%
\vspace{1mm}
\noindent
(8) the McCarthy 91 {function}:  
$M(x)$ $=$ if $x\!>\!100$ then $x\!-\!10$ else $M(M(x\!+\!11))$ 
%%(i) $x\!-\!10$ if $x\!>\!100$ 
%%(iii) $M(M(x\!+\!11))$ otherwise;

\vspace{1mm}
\noindent
(9) the Dijkstra \textit{fusc} function:\nopagebreak

~~~{$\textit{fusc}(0)=0$,} ~~~~$\textit{fusc}(1)=1$, ~~~~$\textit{fusc}(2x)=\textit{fusc}(x)$,

~~~$\textit{fusc}(2x\!+\!1)=\textit{fusc}(x\!+\!1)+\textit{fusc}(x)$, ~~~~~and

%%
%%(i)~{$\textit{fusc}(0)\!=\!0$,} (ii)~$\textit{fusc}(1)\!=\!1$, (iii)~$\textit{fusc}(2x)\!=\!\textit{fusc}(x)$,\\
%%(iv)~$\textit{fusc}(2x\!+\!1)\!=\!\textit{fusc}(x\!+\!1)+\textit{fusc}(x)$, and
%%$F(x)$ is 
%%(i) 1   if $x\!=\!1$, or else 
%%(ii) $F(y)$ if $x\!=\!2\!*\!y$ for some non-negative integer $y$
%%(iii) (ii) $F(y)\!+\!F(y+1)$ if $x\!=\!2\!*\!y\!+\!1$ otherwise;

\vspace{1mm}
\noindent\hangindent=7mm
(10) a function that computes the minimum number of moves needed for the solution of the towers of Hanoi problem.

%%%
\smallskip

Our strategy turns out to be very effective in increasing the ability
of the CHC solver to prove the satisfiability, or the unsatisfiability, of the clauses
encoding the considered problems.
As already mentioned, the main reason for this effectiveness is due to the fact 
that, by pairing together two atoms, 
the Predicate Pairing strategy often enables the discovery of relations between some of their
arguments.

\hangindent=0mm

\subsection{Monotonicity, Injectivity, and Functional Dependence}

Some interesting classes of relational properties we have considered are 
those of monotonicity, injectivity, and functional dependency.
These notions relate
two different terminating executions of the same program on two 
distinct input values, say $x$ and $y$, computing the output
values, say $m$ and $n$, respectively. The definition of
these properties are derived in a straightforward manner
from those of the mathematical functions.

In particular, monotonicity properties state that
the application of the program on ordered input values
produces ordered output values.
For example, a typical monotonicity property is the following: 
if $x\!\leq\! y$,  then  $m\!\leq\! n$.

Injectivity properties state that
any two executions of the same program on different inputs produce different outputs, that is,
if $x\!\neq \!y$,  then  $m\!\neq\! n$.

Functional dependence properties state that the output 
of a program is a function of (a possibly proper subset of) its input values:
for instance, if $x\!=\!y$,  then  $m\!=\!n$.

In particular, 
let us consider the following constrained Horn clauses encoding
the operational semantics of 
a given imperative recursive program {\tt{Fib}} that computes the Fibonacci numbers:

\vspace{.5mm}

$\textit{fib\/}(X,\!Y)  \leftarrow   X\!=\!0, Y\!=\!0$

$\textit{fib\/}(X,\!Y) \leftarrow   X\!=\!1, Y\!=\!1$

$\textit{fib\/}(X,\!Y)  \leftarrow   X\!\geq\!2, X1\!=\!X\!-\!1, X2\!=\!X\!-\!2, 
Y\!=\!Y1\!+\!Y2,\ \textit{fib\/}(X1,\!Y1),\ \textit{fib\/}(X2,\!Y2)$

\vspace{.5mm}
\noindent
where the first argument of \textit{fib\/} encodes the input and the second argument 
of \textit{fib\/} encodes the output.
Then, the above mentioned properties of 
monotonicity, injectivity, and functional dependence
of the program {\tt{Fib}}
can be checked by testing the satisfiability of 
the following clauses:

\vspace{1mm}
\makebox[84mm][l]{$\textit{false}  \leftarrow   Y2\!\geq\!Y1\!+\!1,\ X1\!\geq\!X2,\   \textit{fib\/}(X1,\!Y1),\ \textit{fib\/}(X2,\!Y2)$} \makebox[38mm][r]{(Monotonicity)}

\makebox[84mm][l]{$\textit{false}  \leftarrow   Y1\!=\!Y2, \ {X1\!\neq\!X2},\  \textit{fib\/}(X1,\!Y1),\ \textit{fib\/}(X2,\!Y2)$} \makebox[38mm][r]{(Injectivity)}

\makebox[84mm][l]{$\textit{false} \leftarrow   Y1\!\neq\!Y2,\ X1\!=\!X2,\  \textit{fib\/}(X1,\!Y1),\ \textit{fib\/}(X2,\!Y2)$} \makebox[38mm][r]{(Functional Dependence)}

\vspace{1mm}
\noindent

{Note  that the above clauses are all instances of clause {\it RP} encoding the general 
	relational property.}
	
Based on this example, the reader will not find it difficult to express % alberto
%have no difficulties in expressing 
monotonicity, injectivity, 
and functional dependence for other given imperative programs. 
We have successfully verified these properties
for programs computing: (i)~the sum of two numbers (by iterated increment),
(ii)~the product of two numbers (by iterated addition), and (iii)~the square 
and the cube of a number 
(by iterated addition). We have also considered some more programs 
containing simple, sequential, or nested while-loops,
possibly combined with conditionals.

\subsection{Non-interference}

%We have applied our proof method for verifying that a program satisfies 
Non-interfer\-ence is a property that guarantees information-flow security. It can be viewed as  
a variant of the functional dependence property as we now indicate.

%noninterference stipulates that public outputs of a program should
%be functionally dependent on public input only, and not on private input

Let us consider an imperative program $P$ whose variables  are partitioned into a set of
public variables (or low security variables) and a set of private variables 
(or high security variables).
We say that  
%the program 
$P$ satisfies the non-interference property %security property
if %and only if
any two terminating executions of $P$,
starting with the same initial values of the public variables,
but possibly with different values of the private variables,
compute  %in states (++++configurations? check+++) 
the same values of the public variables.
%, regardless of the values of the private variables.
%
Thus, if a program satisfies the non-interference property,
an attacker cannot acquire information about the private variables by observing 
the input/output relation between the public variables,
which are functionally dependent on the public input variables only.
%(Indeed non-interference is a special type of functional dependence).

To clarify the ideas, let us consider
the following simple imperative program~${\mathtt{HL}}$:

\hspace*{6mm}\begin{minipage}[t]{1\textwidth}
\begin{verbatim}
while (high >= 1) { high = high-1; low = high; }
\end{verbatim}
\end{minipage}

\noindent
where $\texttt{low}$ is a public variable and $\texttt{high}$ is a private variable.
Program ${\mathtt{HL}}$ violates the non-interference property
because there exist
two different executions 
starting with identical values of the variable $\texttt{low}$ and 
terminating in configurations
having different values for variable $\texttt{low}$. Indeed, 
if initially we have that~$\texttt{high}$ is at least~$\texttt{1}$, then the body
of while-loop is executed and the final value of  $\texttt{low}$ will be $\texttt{0}$,
otherwise the value of $\texttt{low}$ is left unchanged.

The non-interference property for program $P$ 
can be verified by checking the satisfiability of 
the following set of clauses:

\vspace{1mm}
$\textit{false} \leftarrow\textit{OutL}\!\neq\!\textit{OutL}1,\ L\!=\!L1,\  p(L,H,\textit{OutL}),\ p(L1,H1,\textit{OutL}1)$ 

$p(L,H,\textit{OutL}) \leftarrow H\!<\!1,\ \textit{OutL}\!=\!L$

%$p(L,H,\textit{OutL}) \leftarrow  H\!\geq\!1,\ H1\!=\!H\!-\!1,\ L1\!=\!H1,\   p(L1,H1,\textit{OutL})$
$p(L,H,\textit{OutL}) \leftarrow  H\!\geq\!1,\ H1\!=\!H\!-\!1,\ L1\!=\!H1,\   p(L1,H1,\textit{OutL})$

\vspace{1mm}

\noindent where: (i)~the predicate $p(L,H,\textit{OutL})$ 
encodes the input/output relation among the variables of program $P$,
(ii)~the variables $L$ and $H$ encode the values of the variables  
$\texttt{low}$ and $\texttt{high}$ 
at the beginning of the while-loop,  and (iii)~$\textit{OutL}$ encodes the value of the variable 
$\texttt{low}$ %upon termination 
at the end of the while-loop.
Note that the set of clauses shown above is \textit{unsatisfiable}
because program $P$ \textit{violates} the non-interference property.

{The reader may note that the first clause, encoding the
non-interference property for program $P$, is an instance of 
clause {\it RP} defining the general relational property.
The encoding of the non-interference property for other programs 
can be done in a similar way.

\smallskip
The following program ${\mathtt{HL1}}$ is representative of a class
of programs  for which we have {successfully} verified that the non-interference property holds:

\noindent
\begin{minipage}[t]{1\textwidth}
\begin{verbatim}
 low1 = low2;   low1 = low1 + f(high);   low1 = low1 - g(high,low1);
\end{verbatim}
\end{minipage}

\noindent
where: (i)~$\texttt{low1}$ and $\texttt{low2}$ are public variables, $\texttt{high}$ is a private variable,
and (ii)~$\texttt{f}$ and $\texttt{g}$ are two 
functions defined as follows:

% NONINT/sumupto2
\smallskip
\begin{minipage}[t]{1\textwidth}
\begin{verbatim}
int f(int m) {
  int i = 0, s = 0;
  while (i <= m) { s += i+m; i++; }
  return s;  }
 \end{verbatim}
\end{minipage}

\begin{minipage}[t]{1\textwidth}
\begin{verbatim}
int g(int m, int n) { 
  int i = 0, s = 0; 
  if (n <= m) { while ( i<= n) { s += i+m; i++; } } ;   
  while (i <= m) { s += i+m; i++; } 
  return s; } 
\end{verbatim}
\end{minipage}

\vspace{2mm}
\noindent
Note that, in the  program~${\mathtt{HL1}}$ the functions $\texttt{f}$ and $\texttt{g}$ 
compute the same value. 
This program~${\mathtt{HL1}}$ \textit{does satisfy}
the non-interference property,
and thus the corresponding set of clauses is \textit{satisfiable}.
Indeed, in the program~${\mathtt{HL1}}$ the public variable~\texttt{low1} is first incremented and then decremented by the same value,
which, however, is computed by the distinct, yet equivalent functions \texttt{f} and \texttt{g}, which take the private variable \texttt{high} as input.

\subsection{Loop Optimizations}

Modern compilers
often perform a series of optimizations for producing a new program 
that is semantically equivalent to an old program, but
whose execution is faster, or requires less memory, or has lower energy consumption.

By applying our method % for checking the equivalence of CHCs 
we have successfully verified equivalence properties
between some imperative programs 
and their optimized versions~\cite{LoM16}.
{The CHC encoding of program equivalence is the one defined by clause {\it EQ} in Section~\ref{subsec:nlin}.}
%\comment{+++ cite Lopez CORK +++}
For instance, we have proved the equivalence of the following program:

\smallskip
\begin{minipage}[t]{1\textwidth}
\begin{verbatim}
while (i < n) {
   if (n > 5) { a = a+n;   i = i+1; } 
      else    { a = a+1;   i = i+1; }
} 
\end{verbatim}
\end{minipage}

\smallskip
\noindent
and the one derived from it by the \textit{loop unswitching} optimization:\nopagebreak

\smallskip
\begin{minipage}[t]{1\textwidth}
\begin{verbatim}
if (n > 5) { while (i < n) { a = a+n;   i = i+1; } } 
   else    { while (i < n) { a = a+1;   i = i+1; } }
\end{verbatim}
\end{minipage}

\smallskip
\noindent
where the conditional statement occurring in the while-loop 
is moved outside the loop, so that
the evaluation of the conditional expression is performed only once, instead
of being performed at each loop iteration.
%%Also, the statements occurring inside the conditional branches of the original program 
%%are looped over independently in the optimized program.
%%
%%%The optimized program produced by the {loop unswitching} transformation
%%%is shown below:
%%\smallskip
%%\begin{minipage}[t]{1\textwidth}
%%\begin{verbatim}
%%if (n > 5) { while (i < n) { a = a+n;   i = i+1; } } 
%%   else    { while (i < n) { a = a+1;   i = i+1; } }
%%\end{verbatim}
%%\end{minipage}

%\vspace{2mm}
%\noindent
We have also considered some specific instances of other equivalence problems
relating original, non-optimized programs
to new programs obtained by applying the following loop optimizations:

\noindent\hangindent=5mm
(i)~\textit{loop fission}, that splits the commands occurring in a loop in two blocks
that are then executed by two consecutive, independent loops;

\noindent\hangindent=5mm
(ii)~\textit{loop fusion}, that merges the commands occurring in consecutive loops and executes them in a single loop;

\noindent\hangindent=5mm
(iii)~\textit{loop reversal}, that executes the commands occurring in a loop, in a new loop where
the iteration proceeds in reversed order
with respect to the order of the original loop;  % spin13 

\noindent\hangindent=5mm
(iv)~\textit{strength reduction}, 
that replaces iterated expensive computations in a loop 
by cheaper ones (for instance, replacing multiplication by a loop index with addition); and
 
\noindent\hangindent=5mm
(v)~\textit{code sinking}, that moves code occurring immediately before or after a loop inside the loop itself,
possibly using conditionals for keeping the semantics of the program unaltered. 

\hangindent=0mm
\noindent
We have also considered other loop optimizations, including {\it{loop tiling}}, {\it{loop aligning}}, {\it{loop pipelining}} as well as other optimizations for removing redundant assignments, expression evaluations, and conditionals. % ex1   % condrem

We are confident that our method of proving equivalence of programs can be 
extended for proving correctness of code optimizations  
at a schematic level~\cite{Ler09,LoM16}, and not for some specific instances only.
We leave this study for future research.

%by introducing extra predicates for modeling generic program fragments
%and their pre-conditions and post-conditions. 
%Verifying the effectiveness of our method and of the CHC solvers at proving the correctness 
%of the optimizations in the general case can be the object of future research.

\subsection{Array-manipulating Programs}

We have applied our verification method to relational properties 
of imperative programs
manipulating integer variables and integer arrays.
%%% code-sinking lavora su array
%%Note, however, that our framework is  able to
%%prove relational equivalence properties between CHCs 
%%with constraints over different domains and 
%%we will now show how to deal with CHCs 
%%derived from programs manipulating integer arrays.

Let us first introduce some preliminary notions.
%% ex Fund INF
An {\it integer array} {\it a}
(or an {\it   array}, for short) is a finite sequence of integers
whose length, called the dimension of the array, is denoted ${\textit{dim\/}}({\it a})$.
An {\it  atomic array constraint} is %an expression of the form 
either
\textit{read\/}$(a,i,v)$, denoting that the~\mbox{$i$-th} 
element of the array~$\mathit{a}$ is the integer~$v$, 
or~\textit{write\/}$(a,i,v,b)$, denoting that, 
for $k\!=\!1,\ldots, {\textit{dim}}({\it a})$, if ${k\!\neq\! i}$, then
the $k$-th element of
{\it a} is equal to the $\textit{k}$-th element of {\it b}, and if ${k\!=\!i}$, then
the $k$-th element of $b$ is the integer~$v$.
%\comment{+++ For more details on we refer the reader to FundInf17?++++.}

The \textit{read} and  \textit{write} constraints satisfy the
following implicative axioms~\cite{Br&06}, 
whose variables are assumed to be universally quantified at the front:

\smallskip
\noindent %\fbox { % beginning of frame
\hspace*{-4mm}\begin{minipage}{1.2\textwidth}
\noindent
\makebox[100mm][l]{\makebox[6mm][l]{}\makebox[41mm][l]{
$\big(I\!=\!J,\ \textit{read\/}(A,I,U),$}      $\textit{read\/}(A,J,V)\big)
~\rightarrow~ U\!=\!V$}({\it array congruence}\/)

\noindent
\makebox[102mm][l]{\makebox[6mm][l]{}\makebox[41mm][l]{
$\big(I\!=\! J,\ \textit{write\/}(A,I,U,B),$}  $\textit{read\/}(B,J,V)\big) 
~\rightarrow~ U\!=\!V$}
({\it read-over-write})

\noindent
\makebox[102mm][l]{\makebox[6mm][l]{}\makebox[41mm][l]{
$\big(I\!\neq\! J,\ \textit{write\/}(A,I,U,B),$} $\textit{read\/}(B,J,V)\big) 
~\rightarrow~ \textit{read\/}(A,J,V)$}
({\it read-over-write})
\end{minipage}
% }  % end of frame

%%%%%%%%%%%%%%%%%%%

\smallskip
\noindent
For example, the operational semantics of the following imperative program
which acts on the array {\tt a}: \nopagebreak

\vspace{1mm}
\begin{minipage}[t]{1\textwidth}
\begin{verbatim}
  i = 1;   a[0] = 3; 
  while ( i < n ) { a[i] = a[i-1]+2;  i++; } 
\end{verbatim}
\end{minipage}

\vspace{2mm}
\noindent 
can be represented by the following set of CHCs:

\vspace{1mm}
$\textit{prog\/}(N,A1,A3) \leftarrow I\!=\!1,\ K\!=\!0,\ U\!=\!3,\  \textit{write\/}(A1,K,U,A2),\  \textit{loop\/}(N,A2,I,A3)$

$\textit{loop\/}(N,A1,I,A3) \leftarrow I+1 \!\leq\! N,\  J\!=\!I\!-\!1,\  \textit{read\/}(A1,J,U), $

$\hspace{20mm} V\!=\!U\!+\!2,\ \textit{write\/}(A1,I,V,A2),\ I1\!=\!I\!+\!1,\ \textit{loop\/}(N,A2,I1,A3)$
		  
$\textit{loop\/}(N,A,I,A) \leftarrow I\!\geq\!N$

\vspace{1mm}
\noindent 
In these clauses: (i)~the predicate $\textit{loop\/}(N,A1,I,A2)$ encodes the while-loop,
(ii)~its arguments $N,A1$, and $I$ encode the values of variables $\tt n,a$, and $\tt i$, respectively,
at loop entry, % when entering the loop
and (iii)~$A2$ encodes the value of \texttt{a} at loop exit.

Now we show an example of an equivalence property between array manipulating programs
that has been proved by using our method based on the Predicate Pairing strategy.
%We have proved the equivalence between the sets of 
%CHCs derived from the two
Let us consider the  
programs~${\mathtt{P1}}$ and~${\mathtt{P2}}$ shown in Table~\ref{looppipelining}, where 
program~${\mathtt{P2}}$ is obtained from program~${\mathtt{P1}}$
by applying the loop-pipelining %compiler 
optimization,
a commonly used technique for enabling
instruction-level parallelism at the hardware level. 
%thereby increasing the number of instructions executed by the CPU per time unit.

\begin{table}[ht]
\begin{center}
\begin{tabular}{c|c}
Program ${\mathtt{P1}}$\hspace{20mm} & Program ${\mathtt{P2}}$\hspace{20mm} \\[-2mm]
\hline\\[-7mm]
\begin{minipage}{0.5\textwidth}
\vspace{3mm}
\begin{verbatim}
   i=0;
   while (i < n) {
      a[i]++;  
      b[i] += a[i];  
      c[i] += b[i];  
      i++;
   }
\end{verbatim}
\end{minipage}
& 
\begin{minipage}{0.5\textwidth}
\vspace{3mm}
\begin{verbatim}
   i = 0; 
   a[0]++; 
   b[0] += a[0]; 
   a[1]++;
   while ( i < n-2 ) {
      a[i+2]++;  
      b[i+1] += a[i+1]; 
      c[i] += b[i]; 
      i++;
    }
   c[i] += b[i]; 
   b[i+1] += a[i+1]; 
   c[i+1] += b[i+1];
\end{verbatim}
\end{minipage}

\end{tabular}
\end{center}
\caption{\rm {The source program ${\mathtt{P1}}$ (left) and the optimized program ${\mathtt{P2}}$ obtained by applying the  loop pipelining transformation (right).\label{looppipelining}}}
\end{table}%

The equivalence between programs ${\mathtt{P1}}$ and ${\mathtt{P2}}$
{with respect to the output array~{\tt c}},
is expressed by the following clause~$F$ {(which is an instance of clause {\it EQ} of Section~\ref{subsec:nlin})}: \nopagebreak

$F\/$:~ $\textit{false} \leftarrow X\neq Y,\ N\geq1,\ J\geq0,\ J\leq N\!-\!1,\ 
  \textit{read\/}(C1,J,X), \textit{read\/}(C2,J,Y),$\nopagebreak

\hspace{13mm} $\clspace \textit{new\/}11(N,A,C1),\ \textit{new\/}21(N,A,C2) $

\noindent
where~$\textit{new\/}11(N,A,C1)$ represents the input/output relation of the 
source program~${\tt P1}$,
and in particular, $N$ is the value of the integer variable {\tt n}, 
$A$  is the value of the array {\tt a} at the beginning of program execution, 
and $C1$ is the value of the array {\tt c} at the end of program execution.
Similarly, $\textit{new\/}21$ represents the input/output relation of the 
optimized program~${\tt P2}$.

Thus, by proving the satisfiability of the set of clauses
consisting of $F$ together with the clauses 
defining $\textit{new\/}11$ and~$\textit{new\/}21$,
we have been able to prove that programs ${\mathtt{P1}}$ and~${\mathtt{P2}}$
produce identical values for the array {\tt c} as output,
when provided with identical values for the array {\tt a} as input.

\section{Experimental Evaluation}
\label{sec:experiments}
\newcommand{\verimap}{\textsc{VeriMAP}\xspace}
\newcommand{\smtchk}{$\mathit{HSC}$\xspace}
\newcommand{\eldarica}{\textsc{Eldarica}\xspace}

\newcommand{\ite}{{\sc ite}\xspace}
\newcommand{\rec}{{\sc rec}\xspace}
\newcommand{\ir}{{\sc i-r}\xspace}
\newcommand{\arr}{{\sc arr}\xspace}
\newcommand{\mon}{{\sc mon}\xspace}
\newcommand{\inj}{{\sc inj}\xspace}
\newcommand{\fun}{{\sc fun}\xspace}
\newcommand{\lff}{{\sc comp}\xspace} 
\newcommand{\nlin}{{\sc nlin}\xspace}
\newcommand{\spec}{{\sc pcor}\xspace}
\newcommand{\lopt}{{\sc lopt}\xspace}
\newcommand{\noint}{{\sc nint}\xspace}

In this section we present the experimental evaluation we have performed
for assessing the effectiveness of the Predicate Pairing  
strategy (PP strategy, for short) presented in Section~\ref{sec:strategy}. 

\vspace*{-3mm}
%===============================================================================
\paragraph{Implementation.} 
We have implemented Algorithm~\ref{alg:pairing1} by using the \verimap
system~\cite{De&14b}, which is a tool for software model checking based on  
transformation techniques for CHCs. 
Then we have used the SMT solver~\textsc{Z3} \cite{DeB08} for checking 
the satisfiability of the clauses generated by \mbox{\verimap}. 
In particular, we have used Z3 version 4.5.0 with the Duality fixed-point 
engine \cite{McR13}, which provides support for constraints defined
on linear integers and integer arrays. 
%\comment{DELETE?: This version of Z3 is required to deal with the verification problems
%we have considered in our experimental evaluation.}

Our prototype implementation consists of two components: (1)~a module that 
realizes Algorithm~\ref{alg:pairing1}, and (2)~a module that translates 
the generated CHCs into the SMT-LIB format which is the format accepted by Z3
(see Figure~\ref{fig:system}). %Duality. 
The VeriMAP system
also provides a front-end module~(T) that takes a pair of C programs, 
together with a relational property to be verified, 
and translates them into the CHCs 
that encode the verification problem~\cite{De&16c}.

%%%%%%%%%%%%%%%%==============================================================
%% Alberto use TeXShop with: tex and dvi. 
%% The numbers for the bounding box bb is given by the translator: pdf2ps
%% starting from the pdf.
%\begin{figure}[ht]
%\vspace{3mm}
%\centering
%\includegraphics[width=1\linewidth,bb=116 597 474 687]{Fig-Transf-System.pdf}
%\caption{The Verification System: (1)~the Predicate Pairing module, implementing 
%Algorithm~\ref{alg:pairing1}, and (2)~the Translator to SMT-LIB module. \label{fig:system}}
%\vspace{2mm}
%\end{figure} 

\begin{figure}[ht]
	\vspace{3mm}
	\centering
	\includegraphics[scale=1]{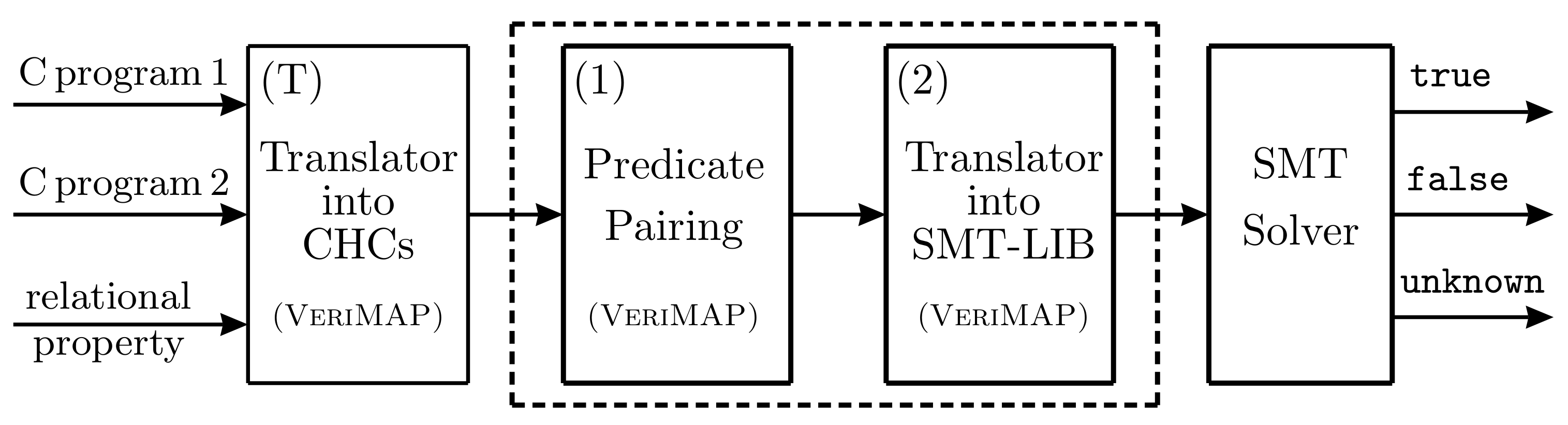}
	\caption{The Verification System: (1)~the Predicate Pairing module, implementing 
		Algorithm~\ref{alg:pairing1}, and (2)~the Translator to SMT-LIB module. \label{fig:system}}
	\vspace{2mm}
\end{figure}

\vspace*{-3mm}
%===============================================================================
\paragraph{Verification problems.}
We have considered a benchmark\footnote{The benchmark set can be found at \url{http://map.uniroma2.it/predicate-pairing}} 
of 163 sets of CHCs (153~of which are satisfiable 
and the other~10 are unsatisfiable), representing verification problems 
{(all acting on integers and 14 of them acting on integer arrays)}, 
which refer to relational properties of small, yet non-trivial, imperative programs 
mostly taken from the literature~\cite{Ba&11,Ben04,Fe&14,De&15c,De&16c,LoM16}.

%
%All programs act on integers, except 
%(i)~all programs in the \arr category,
%(ii)~7 out of~10 in the \lff category, and
%(iii)~1 out 20 in the \lopt which act on integer arrays.%
%%Besides the properties presented in Section~\ref{sec:case_studies}, that is,
%%monotonicity (\mon), injectivity (\inj), functional dependence (\fun), 
%%loop optimizations (\lopt), and non-interference (\noint), 

We have considered the following categories of relational verification 
problems.
(1)~The category \nlin, which refers to equivalence properties between  
programs implementing functions with nonlinear and/or nested recursion. 
For instance, we have verified the equivalence of two
programs for computing the Ackermann function (see 
the running example presented in Section~\ref{sec:strategy} and the examples of
Section~\ref{subsec:nlin}). 
(2)--(6)~The categories \mon, \inj, \fun, \noint, and \lopt, which refer to
monotonicity, injectivity, functional dependence, 
non-interference, and loop optimizations problems, respectively
 (see Section~\ref{sec:case_studies}).
(7)~The category \ite, which refers to
equivalence properties and inequality properties relating two iterative programs
acting on integers (by an inequality property we mean that the values computed by the
programs are related by~$\leq$).
(8)~The category \arr, which refers to equivalence and inequality properties 
between iterative programs acting on integer arrays. 
(9)~The category \rec, which refers to equivalence properties between
recursive programs.
(10)~The category \ir, which refers to equivalence properties between
an iterative program and a (non-tail) recursive program.
For example, we have verified the equivalence of the iterative and recursive versions 
of programs for computing: (i) the greatest common divisor of two integers, and 
(ii)~the $m$-th triangular number, that is, $\sum_{x=1}^{m}x$.
%$1+2+\ldots+m$    %$T_m\!=_{\small{def}}\!\sum_{i=1}^{m}i$.
%
(11)~The category \lff, which refers to equivalence and inequality properties 
between programs that contain compositions of 
different numbers of loops acting on integers (3~problems) and integer arrays 
(7~problems).
(12)~The category~\spec, which refers to partial correctness properties of
an iterative program with respect to a recursive functional postcondition~\cite{De&15c}.

Note that in our benchmark set, 31 problems (belonging to the categories \nlin,  \noint, and \lff) 
are encoded by sets of clauses that include nonlinear clauses, besides the ones that encode the
relational properties, which are always nonlinear (see Sections \ref{sec:strategy} and ~\ref{sec:case_studies} for some examples). 

%MAU: non chiara la distinzione tra le clausole che codificano la property
%e quelle che codificano il problema}
%
%\comment{NOTA: "nonlinear programs" viene usato nella descrizione della categoria NLIN}

%OPPURE: refer to nonlinear programs.

%By excluding the property, we
%have 31 sets of CHCs including nonlinear clauses (belonging to the categories comp,
%nlin, and nint).

\vspace*{-3mm}
%===============================================================================
\paragraph{Technical resources.}
The experimental evaluation has been performed on a single core of an Intel Core 
Duo E7300 2.66GHz processor with 4GB of memory running Ubuntu. 
For all problems we have set the timeout limit of 300 seconds.

\vspace*{-3mm}
%=============================================================================== 
\paragraph{Experimental processes.}
We have considered the following two experimental processes.
(E1)~The first experimental process consists in running Z3 for checking the satisfiability 
of the original sets of CHCs that encode the verification problems. (E2)~The second 
experimental process consists in running Algorithm~\ref{alg:pairing1} on the 
original sets of CHCs for each verification problem, and then running Z3 for checking 
the satisfiability of the derived CHCs. In this second process
the PP strategy has been iterated~(see the end of Section~\ref{sec:strategy}), 
if more than one pair of atoms
was present in the bodies of the original sets of clauses.
In particular, PP has been iterated for 24 sets of CHCs, in total,
belonging to categories  \noint, \arr, \lff,  and  \spec.
% for 1 set of CHCs in the \arr category and 13 
%sets of CHCs in the \spec category.

%===============================================================================
% EXPERIMENTAL RESULTS
\begin{table}[ht]
\begin{center}
\begin{tabular}{@{\hspace{2pt}}r@{\hspace{0pt}}l@{\hspace{0pt}}r@{\hspace{14pt}}| %id,category,P
@{\hspace{0pt}}r@{\hspace{8pt}}r@{\hspace{14pt}}|                %S1,T1
@{\hspace{0pt}}r@{\hspace{8pt}}|                                %PP
@{\hspace{0pt}}r@{\hspace{8pt}}r@{\hspace{14pt}}                 %S2,T2
}            
\multicolumn{3}{c|}{\textrm{\textrm{Problems~~}}}   &  %id,category,P
\multicolumn{2}{c|}{\textrm{Z3} before \textrm{PP}~~~~~~}           &  %S1,T1
\multicolumn{1}{c|}{\textrm{PP}~}                   &  %PP,PP on S2
\multicolumn{2}{c}{\textrm{Z3} after \textrm{PP}~~~~~~}       \\[-2mm] \hline %S2,T2
%%%
\multicolumn{2}{l}{\textrm{Category}}  & $P_{_{_{_{_{}}}}}$ & $S_1$ & $T_1$~~ & 
$T_{\rm PP}$
&$S_2$ & $T_2$~~ \\[-2mm]\hline\hline

%%%
(1)  & \nlin  & 13  &  4 &  16.11  & 25.80 & 13  &  13.12 \\[-2mm]\hline
%%%
(2)  & \mon   & 18  &  1 &   1.04  &  2.27 & 12  &   3.72 \\[-2mm]\hline
%%%
(3)  & \inj   & 11  &  0 &     --  &  1.36 &  8  &   1.39 \\[-2mm]\hline
%%%
(4)  & \fun   &  7  &  4 &   1.39  &  1.24 &  7  &   1.48 \\[-2mm]\hline
%%%
(5)  & \noint &18   & 3  & 0.27    & 55.80 & 17  &  41.33 \\[-2mm]\hline
%%%
(6)  & \lopt   & 20  &  2 &   4.83  &  2.98 & 15  &  10.71 \\[-2mm]\hline
%%%
(7)  & \ite   & 22  &  5 &  26.67  &  4.53 & 18  &  17.01 \\[-2mm]\hline
%%%
(8)  & \arr   &  6  &  1 &   7.45  &  2.04 &  5  &   3.25 \\[-2mm]\hline
%%%
(9)  & \rec   & 15  &  6 &   2.89  &  1.50 & 13  &   4.28 \\[-2mm]\hline
%%%
(10)  & \ir    &  4  &  0 &     --  &  0.65 &  3  &   1.02 \\[-2mm]\hline
%%%
(11)  & \lff   & 10  &  0 &   --    & 16.35 &  7  &   6.46 \\[-2mm]\hline
%%%
(12)  & \spec  & 19  &  5 &  83.93  & 17.84 & 17  &  17.65 \\[-2mm]\hline

\multicolumn{2}{l}{\textrm{Total number}}     & 163    & 31 & 144.58 & 132.36 & 135 & 121.42 \\[1mm]
\multicolumn{2}{c}{\textrm{Average Time}}     &  &  &   4.66  &  0.81  &     &   0.90 \\[-2mm]\hline\hline
\end{tabular}
\caption{{\rm{
The first two columns report the names of the problem categories and the number $P$
of problems in each category, respectively. 
Columns $S_1$ and $S_2$ report the number of verification problems solved by Z3 
before and after the application of the PP strategy, respectively.
Columns $T_1$ and $T_2$ report the time taken by Z3 to solve the problems
reported in columns $S_1$ and $S_2$, respectively.
Column~$T_{\rm PP}$ reports the time taken by \verimap to apply the 
PP strategy on the~$P$ problems. Times are in seconds. 
The timeout is of $300$ seconds. No timeout occurred during
the application of the PP strategy.
}}
\label{tab:evaluation}	
}
\end{center}
\end{table}

\vspace*{-3mm}
%===============================================================================
\paragraph{Results.}
The results of the experimental evaluation are summarized in 
Table~\ref{tab:evaluation}. 
The times reported are the CPU seconds spent in user mode and kernel 
mode by \linebreak (i)~\verimap for transforming the clauses, and (ii)~Z3 for checking their 
satisfiability.
The times required for translating the CHCs into the SMT-LIB format (this translation 
is necessary for running Z3) are: 6.39~seconds for the first experimental process~E1, and 
22.59~seconds for the second experimental process~E2. 
Those translation times are not very high %negligible 
with respect to the solving times which are 144.58~seconds and 121.42~seconds, respectively.
%\comment{DELETE?:Moreover, those translation times are irrelevant for evaluating the increasing  
%of the solving capabilities of Z3 due to the (possibly, iterated)
%application of the PP strategy.}
%for evaluating the improvement of 
%the solving performance of Z3. 
%%\comment{
%%Moreover, These translation times are negligible with respect to the solving 
%%times, 6.39 and 18.58 seconds to translate all clauses before and after the 
%%application of the PP strategy, respectively.)
%%}

As expected, the use of the PP strategy significantly increases the number of 
problems that are solved by Z3. 
In particular, the number of solved problems %that are solved by~Z3 
increases from~31 
(Total of Column~$S_1$) to 135 (Total of Column~$S_2$). Note, however,
the PP strategy can derive a set of clauses which is larger than the 
original set. In our benchmark we have observed that size increases of about 
2.16 times on average.

%%The set of CHCs produced as output by the PP strategy can be larger than the 
%%set of CHCs provided as input.
%%\comment{In our benchmark we have observed that the size increases of about 
%%2.16 times on average.}
%%However, despite the increase of size, the PP strategy significantly improves 
%%the efficacy of the CHC solver. 

Note also that the PP strategy increases the efficiency of the 
satisfiability check.
Indeed, the average time taken to run PP and then Z3 
{(1.88 seconds)} is lower than the average time taken to run Z3 on 
the original set of clauses {(4.66 seconds)}.
Although we have proved that the application of the PP strategy preserves
all LIA-definable models, in our experiments
we have found three problems which Z3 was able to solve 
before the application of the~PP strategy (i.e., in the first 
experimental process), and it was no longer able to solve, within the given time limit, after
the application of the~PP strategy
(i.e, in the second 
experimental process). 
This phenomenon may be due to the
fact that the termination of the algorithms implemented by the Z3 solver is
sensitive to the syntactic form of the input clauses, and the PP strategy modifies that form.
Further work is needed to improve the termination behavior of the solver.

Finally, we would like to point out that the application of the
PP strategy does not decrease the
efficiency of the whole verification process. If we consider only the 28 problems
for which Z3 is able to solve both before and after the application 
of the PP strategy, the average time taken to run PP and then Z3 (1.58 seconds) 
is slightly lower than the time taken by Z3 alone (1.77 seconds).

\section{Related Work  and Conclusions}
\label{sec:related}
The basic idea behind the Predicate Pairing  transformation strategy
for constrained Horn clauses is
that, by finding a recursive definition of a predicate denoting
the conjunction of two atoms, it is often possible to infer relations
among the variables occurring in the two atoms which would have been impossible to discover
by considering each atom separately.

Techniques for transforming logic programs by deriving new predicates defined 
in terms of conjunctions of atoms have been largely studied.
Let us recall, for instance, the well-known
{\it tupling} transformation 
strategy~\cite{PeP94} and the {\it conjunctive partial deduction} 
technique for logic program specialization~\cite{De&99}. 
The main objective of tupling and conjunctive partial deduction
is the derivation of more  efficient 
logic programs by avoiding multiple traversals of data structures and 
repeated evaluations of predicate calls, and by
producing specialized program versions. 

{Thus, Predicate Pairing shares with tupling and conjunctive partial deduction
the  idea of promoting a conjunction of atoms to a new predicate.
However, in this paper we have shown that the application of this idea
to constrained Horn clauses can also play a key role in improving 
the effectiveness of CHC solvers for proving properties
of imperative programs, besides the optimization of the execution of 
logic programs, which is the objective of tupling and conjunctive partial deduction.} 
This is the case especially
when the CHC solvers are required to test the satisfiability of clauses
that encode relational program properties, that is, properties 
that relate two programs, or two executions of the same program.
Indeed, as shown by many examples considered in this paper, state-of-the-art solvers
often fail to prove the satisfiability of sets of clauses
encoding relational properties
because they can only infer relations among the variables of individual atoms.

We have considered CHC solvers that prove satisfiability by using
predicate abstraction, that is, by looking 
for  models that are definable in a specific class $\mathcal A$ 
of constraints~\cite{Bj&15}.
We have shown that, in principle, the Predicate Pairing strategy cannot
worsen the effectiveness of the CHC solver. Indeed, we have proved
a very general result concerning the unfold/fold transformation
rules used by the strategy: if a set of clauses is transformed
by applying the \mbox{unfold/fold} rules, and the original set of 
clauses has an $\mathcal A$-definable model, 
then also the transformed set of clauses has an $\mathcal A$-definable model.
{Thus, if the CHC solver is able to find an $\mathcal A$-definable model
whenever it exists (and this is indeed possible if the validity problem for the 
constraints in $\mathcal A$ is decidable), then every set of clauses that 
can be proved satisfiable by the solver 
before the transformation, will also be proved satisfiable by the solver after the transformation.
We have shown that, in practice, for the Z3 solver this property is
guaranteed with very few exceptions.}

We have also given some restrictions on the use of the rules that guarantee
that the converse of the above property holds, that is:
if a set of clauses is transformed
by applying the unfold/fold rules, and the transformed set of 
clauses has an \mbox{$\mathcal A$-definable} model, 
then also the original set of clauses has an $\mathcal A$-definable model.
However, this property is not always desirable. Indeed, the fact that in some cases
Predicate Pairing is able to transform (satisfiable) clauses that
do not have an $\mathcal A$-definable model into  
clauses that have an $\mathcal A$-definable model, may be a great
advantage. Indeed, this means that while a CHC solver that
looks for  $\mathcal A$-definable models is not able to prove
the satisfiability of the original clauses, the same solver may be
able  to prove the satisfiability of the transformed clauses.

{The study of the properties that relate unfold/fold transformations and 
the existence of $\mathcal A$-definable models 
is not present in the literature on
tupling and conjunctive partial deduction.}

Then, we have presented an algorithm that implements the
Predicate Pairing strategy. One of the novel points
of this algorithm {with respect to tupling and conjunctive
partial deduction}
is that it realizes a heuristic to choose the
appropriate  atoms to be paired together in a new
predicate definition, by maximizing the number of equality constraints
that relate the variables occurring in a pair of atoms.
This heuristic is crucially needed when dealing with nonlinear clauses
that, by unfolding, may generate clauses with more
than two atoms in their body.
We have implemented our algorithm on the VeriMAP 
transformation and verification
system~\cite{De&14b}, 
and we have evaluated its effectiveness on a benchmark of over
160 problems encoding relational properties of small,
yet nontrivial C-like programs. The  properties were of various
kinds, including equivalence, injectivity, functional dependence, 
and non-interference (a property of interest for 
enforcing software security). The results show that the use of
Predicate Pairing as a preprocessor greatly improves
the ability of the Z3 CHC solver
(with the Duality fixed-point computation engine~\cite{McR13}) to prove satisfiability.

Several transformation-based techniques for constrained Horn clauses,
or constraint logic programs, have been proposed as a means of {facilitating}
program verification. 
Many of them are non-conjunctive specialization techniques, which work by
propagating the constraints occurring in the goal, thereby producing
clauses with strengthened constraints in their bodies~\cite{Al&07,De&14c,De&17b,De&15c,KaG15a,KaG17,Me&07,Pe&98}. 
Even if specialization techniques have been shown to be very successful,
in most cases they cannot achieve the same effect as Predicate Pairing. 
Indeed, as already mentioned, Predicate Pairing works by introducing new 
predicates corresponding
to conjunctions of old predicates, whereas non-conjunctive specialization
can only introduce new predicates that correspond to instances of old 
predicates. We have experimentally checked that most
of the problems considered in Section~\ref{sec:experiments}
cannot be solved via {(non-conjunctive)} 
specialization alone. Due to lack of space we have not reported these results.

The query-answer transformation (and variants thereof)
is another pre-processing technique that is sometimes applied
before performing satisfiability tests using CHC solvers~\cite{De&14c,KaG15a,KaG17}.
The aim of this transformation is to simulate the top-down,
goal oriented evaluation of the clauses in a bottom-up framework.
The results we have presented here, showing that Predicate Pairing
is able to transform clauses without an \mbox{$\mathcal A$-definable} model
into clauses with an $\mathcal A$-definable model, are independent of the
evaluation strategy adopted by the CHC solvers, and hence the
query-answer transformation and the Predicate Pairing strategy should be viewed
as orthogonal techniques.

Predicate Pairing is an extension of the {\it Linearization} transformation,
whose objective is to transform a set of linear clauses (that is, clauses
with at most one atom in their body) together with a nonlinear goal, into 
a set of linear clauses and linear goals~\cite{De&15d}.
The Predicate Pairing strategy does not need any linearity assumption, and indeed
in Sections~\ref{sec:strategy}, \ref{sec:case_studies}, 
and~\ref{sec:experiments} we have shown that this strategy can solve
several verification problems encoded by sets of nonlinear 
clauses. 
It has also been shown that Linearization
preserves the existence of LIA-definable models~\cite{De&15d}.
Here we have generalized this result by proving that the application
of the unfold/fold transformation rules, independently of the strategy,
preserves the existence of $\mathcal A$-definable models, for any class of constraints.

The Predicate Pairing algorithm presented here is an improvement of the
one reported in previous work presented at the SAS Symposium~\cite{De&16c}.
Indeed, as already mentioned, here we use an equality-based heuristic
to choose the appropriate  atoms to be paired together, and this technique 
has been shown very effective in practice for handling nonlinear clauses.
Also the case studies and the benchmark set we consider in the present paper are much larger,
and include verification problems such as non-interference, correctness
of loop optimizations, and equivalence of nonlinear recursive programs
that have not been considered in our SAS paper~\cite{De&16c}.
Neither in that paper there are general results 
concerning the preservation of $\mathcal A$-models. 

Bj{\o}rner et al. have shown that  unfolding
preserves $\mathcal A$-definable models provided
that the set $\mathcal A$ of constraints admits Craig interpolation~\cite{Bj&15}.
In the present paper we have generalized this result by considering
also other transformations, and in particular folding, and
we dropped the assumption about Craig interpolation.
Moreover, we assume that   $\mathcal A$ is a subset of the
set $\mathcal C$ of constraints over which the clauses are
defined, while Bj{\o}rner et al.  take $\mathcal A$ to be equal to $\mathcal C$.
Our generalization is significant because sometimes
CHC solvers that make use of predicate abstraction look
for models defined in subsets of the constraints used for the clauses, 
such as the popular domain of the octagons~\cite{Min06}. 

The problem of verifying relational properties is very relevant
in the context of software engineering.
Indeed, during software development it is often the case that one 
modifies the program text, and hence needs a proof that
the semantics of the new
program version has some specified relation to the semantics 
of the old version. This kind of proof is sometimes called 
{\it regression verification}~\cite{GoS08}.

Several logics and methods have been presented in the literature for reasoning
about various relational program properties.
A Hoare-like axiomatization of relational reasoning for simple 
{while} programs has been proposed by Benton~\cite{Ben04},
who however does not present any technique for the automation of the proofs.

{\it Program equivalence} is one of the relational properties that
have been extensively studied in the past, and still receives remarkable
attention in recent work~\cite{Ba&11,Ch&12,Ci&14,Fe&16,Fe&14,GoS08,LoM16,StV16,Ve&12,ZaP08}.
A fruitful idea for easing the problem
of proving  program equivalence is to reduce it to a standard verification
task by using some {\it composition} operator between imperative programs~\cite{Ba&11,Lah&13,ZaP08}.
The application of these operators requires human ingenuity,
and it is still necessary to provide the suitable invariants 
to be used by the program verifier.

A method for reusing available verification techniques to prove 
program equivalence is proposed by Ganty et al.~\cite{Ga&13}, who identify 
a class of recursive programs for which it is possible to precisely compute 
{the so called summaries}. 
This method can be used to reduce the problem of checking the equivalence of two 
recursive programs to the problem of checking the equivalence of their summaries.

Lopes and Monteiro proposed {a different} method for proving program equivalence
that is based on the computation of precise (that is, not
over-approximated) summaries~\cite{LoM16}.  
This method considers programs over the integers
and is based on a transformation into integer (possibly nonlinear) polynomials. 
The equivalence checking algorithm
then works on loop-free programs.
This method has been applied to prove the correctness
of several loop optimizations.
As shown in Section~\ref{sec:case_studies}, Predicate Pairing is
able to prove similar correctness properties, by avoiding the use
of nonlinear arithmetic.

Felsing et al. propose a technique for proving relational properties of
imperative programs, which
is based on a translation of special purpose proof rules
into constrained Horn clauses~\cite{Fe&14}.
The satisfiability of these clauses is then checked by 
state-of-the-art CHC solvers.
The main difference between our approach and the one of Felsing 
et al.~is that we generate a %(different) 
translation of the relational properties
into {sets of CHCs} starting from the semantics of the imperative language, and hence we do not need   
any special 
purpose proof rule that depends on the programming language and the class of 
properties under consideration.
Instead, we use language independent transformation rules {for CHCs}.
In particular, unlike Felsing et al., by using our approach we are 
able to verify relations between programs 
that have different structure,  because the transformation 
rules are independent of the syntax of the source programs. 
	
{In conclusion, we would like to stress that
our work confirms once again the great advantages offered by the 
program verification approach 
based on the use of constrained Horn clauses.
Indeed, by reducing the problem
of verifying properties of programs in a given language
to the problem of reasoning with constrained Horn clauses, 
we are able to use general purpose techniques and very
effective tools
developed over the last four decades in the fields of logic programming,
constraint-based reasoning, and automated theorem proving.}
In this way, we get verification methods with a very high level of  
flexibility and parametricity with respect to the 
language in which programs are written.

\section{Acknowledgments}

We warmly thank the anonymous referees for their very helpful comments and criticism.
This work has been partially supported by the National Group of 
Computing Science (GNCS-INDAM).
E. De~Angelis, F. Fioravanti, and A. Pettorossi are research associates at CNR-IASI, Rome, Italy.

%\bibliographystyle{acmtrans}
%\bibliography{Smc,Transformation} 

\newpage
\section*{Appendix}

\subsection*{Proof of Theorem \ref{thm:tight}}
\begin{proof}
Let us assume that there exists an $\mathcal A$-definable model $\Sigma$ of $P_i$ that is
tight on $\textit{Defs}_i$. 
We will construct an $\mathcal A$-definable model $\Sigma'$ of
$P_{i+1}$ that is tight on $\textit{Defs}_{i+1}$.
The proof proceeds by cases on the transformation rule applied to derive
$P_{i+1}$ from $P_i$.

\medskip
\noindent
({\it Case} R1) Suppose that $P_{i+1}$ is derived from $P_i$ by applying the
definition rule. Thus,  $P_{i+1}=P_i\cup \{D\}$ and
$\textit{Defs}_{i+1}=\textit{Defs}_i \cup \{D\}$, where $D$
is the clause $\textit{newp}(X_1,\ldots,X_k)\leftarrow c,G$, 
and the following conditions hold:
(i)~\textit{newp} is a new predicate symbol,
(ii)~$c \in \mathcal A$, 
(iii)~all predicates occurring in $G$ also occur in $P_0$, and 
(iv)~$X_1,\ldots,X_k$ are distinct variables occurring free in $(c,G)$.

Let $\Sigma'$ be a symbolic interpretation that is equal to $\Sigma$
for all atoms whose predicate is different from \textit{newp}, and
$\Sigma'(\textit{newp}(X_1,\ldots,X_k)) = \exists Y_1\ldots\exists Y_m (c \wedge \Sigma(G)$)
\noindent
where $\{Y_1,\ldots,Y_m\} = \textit{Fvars}(c \wedge \Sigma(G)) \setminus \{X_1,\ldots,X_k\}$. 

Now we have that $\Sigma'$ is an $\mathcal A$-definable model of $P_{i+1}$, as the following two points hold:

\noindent\hangindent=6mm
Point~(i):~$\Sigma'$ is an $\mathcal A$-definable model of $P_{i}$ because $\Sigma'$ is
equal to  $\Sigma$ for all atoms whose predicates occur in $P_{i}$, and

\noindent\hangindent=6mm
Point~(ii):~$\Sigma'$ is an $\mathcal A$-definable model of $D$, that is,
\\
$\mathbb D\models\forall(c \wedge \Sigma'(G) \rightarrow 
\Sigma'(\textit{newp}(X_1,\ldots,X_k)))$.

\hangindent=0mm
Point~(ii) is shown as follows.
Since $Y_1,\ldots, Y_m$ do not occur free in the formula
$\Sigma'(\textit{newp}(X_1,\ldots,X_k))$ and $\textit{newp}$ 
does not occur in $G$, 

$\mathbb D\models\forall(c \wedge \Sigma'(G) \rightarrow 
\Sigma'(\textit{newp}(X_1,\ldots,X_k)))$ 

iff $\mathbb D\models\forall(\exists Y_1\ldots\exists Y_m (c \wedge  
\Sigma(G)) \rightarrow \Sigma'(\textit{newp}(X_1,\ldots,X_k)))$

\noindent
and the latter implication holds by the definition of $\Sigma'$.
Moreover, from the definition of $\Sigma'$ and from the
hypothesis that  $\Sigma$ is tight on $\textit{Defs}_i$, it follows immediately
that $\Sigma'$ is tight on $\textit{Defs}_{i+1}$. 

\medskip
\noindent
({\it Case} R2) Suppose that $P_{i+1}$ is derived from $P_i$ by applying the
unfolding rule. Thus, $P_{i+1}=(P_i \setminus \{C\}) \cup 
\{H\leftarrow  c, {c}_j,G_1, B_j, G_2 \mid  j\!=\!1, \ldots, m\}$, where $C$ is
the clause $H\leftarrow c,G_1,p(X_1,\ldots,X_k),G_2$  in $P_i$
and $\{p(X_1,\ldots,X_k)\leftarrow {c}_j, B_j \mid  j\!=\!1, \ldots, m\}$ 
is the set of clauses in~$P_i$ whose head predicate is $p$.
% and, for $j=1, \ldots, m,$ $c\wedge c_j$ is $\mathbb D$-satisfiable.

Now we show that $\Sigma$ is an $\mathcal A$-definable model of 
$P_{i+1}$ that is tight on $\textit{Defs}_{i+1}$.
By the hypothesis that $\Sigma$ is an $\mathcal A$-definable model of
$P_i$ we have that

$\mathbb D\models\forall(c \wedge \Sigma(G_1) \wedge \Sigma(p(X_1,\ldots,X_k))
\wedge \Sigma(G_2) \rightarrow \Sigma(H))$

\noindent
and, for $j\!=\!1, \ldots, m,$

$\mathbb D\models\forall(c_j \wedge \Sigma(B_j) \rightarrow \Sigma(p(X_1,\ldots,X_k)))$.

\noindent
Then, for $j\!=\!1, \ldots, m,$

$\mathbb D\models\forall(c \wedge c_j \wedge \Sigma(G_1) \wedge \Sigma(B_j)
\wedge \Sigma(G_2) \rightarrow \Sigma(H))$

\noindent
and hence $\Sigma$ is an $\mathcal A$-definable model of $P_{i+1}$.

Obviously, $\Sigma$ is tight on $\textit{Defs}_{i+1}$, which is equal to 
$\textit{Defs}_i$.

\medskip
\noindent
({\it Case} R3) Suppose that $P_{i+1}$ is derived from $P_i$ by applying the
folding rule. Thus, \( P_{i+1}=(P_{i}\setminus\{C\})\cup \{E \} \), where
$C$ is the clause $H\leftarrow c, G_1,Q,G_2$ in $P_i$ and
$E$ is the clause \( H\leftarrow e, G_1, K\vartheta, G_{2} \) derived
%$E$ is the clause \( H\leftarrow c, G_1, K\vartheta, G_{2} \) derived
using  the clause $D$: $K \leftarrow d, B$ in $\textit{Defs}_i$
according to rule R3. Moreover, Conditions
(i)--(iii) listed above when introducing rule~R3, do hold.

Now we show that $\Sigma$ is an $\mathcal A$-definable model of 
$P_{i+1}$ that is tight on $\textit{Defs}_{i+1}$.
By the hypothesis that $\Sigma$ is an $\mathcal A$-definable model of
$P_i$ we have that

$\mathbb D\models\forall(c \wedge \Sigma(G_1) \wedge \Sigma(Q)
\wedge \Sigma(G_2) \rightarrow \Sigma(H))$.

\noindent
By Conditions (ii) and (iii) and the definition of symbolic interpretation, we get that

$\mathbb D\models\forall(e \wedge \Sigma(G_1) \wedge 
(\exists Y_1\ldots \exists Y_m (d \wedge \Sigma(B)))\vartheta
\wedge \Sigma(G_2) \rightarrow \Sigma(H))$

%$\mathbb D\models\forall(c \wedge \Sigma(G_1) \wedge 
%(\exists Y_1\ldots \exists Y_m (d \wedge \Sigma(B)))\vartheta
%\wedge \Sigma(G_2) \rightarrow \Sigma(H))$

\noindent
where $\{Y_1,\ldots, Y_m\} = \textit{Fvars}(d \wedge \Sigma(B))\setminus\textit{Fvars}(\Sigma(K))$.
Since $\Sigma$ is a symbolic interpretation that is tight on $\textit{Defs}_i$, we have that

$\mathbb D\models\forall(e \wedge \Sigma(G_1) \wedge 
\Sigma(K\vartheta)
\wedge \Sigma(G_2) \rightarrow \Sigma(H))$.

%$\mathbb D\models\forall(c \wedge \Sigma(G_1) \wedge 
%\Sigma(K\vartheta)
%\wedge \Sigma(G_2) \rightarrow \Sigma(H))$.

\noindent
Thus, $\Sigma$ is an $\mathcal A$-definable model of 
$P_{i+1}$. 

Obviously, 
$\Sigma$ is tight on $\textit{Defs}_{i+1}$, which is equal to 
$\textit{Defs}_i$.

\medskip
\noindent
({\it Case} R4) Suppose that $P_{i+1}$ is derived from $P_i$ by applying the
constraint replacement rule. Thus, $P_{i+1} = (P_i \setminus \{(H\leftarrow c_1, G), \ldots,(H\leftarrow c_k, G)\}) \cup \{(H\leftarrow d_1, G),$ $\ldots,(H\leftarrow d_m, G)\}$, where

$\mathbb D \models \forall \, (\exists Y_1\ldots\exists Y_r\ (c_1 \vee \ldots \vee c_k) \leftrightarrow \exists Z_1\ldots\exists Z_s\ (d_1 \vee \ldots \vee d_m))$,

\noindent
$\{Y_1,\ldots,Y_r\}=Fvars(c_1 \vee \ldots \vee c_k)\setminus vars(\{H,G\})$,
and $\{Z_1,\ldots,Z_s\}=Fvars(d_1 \vee \ldots \vee d_m)\setminus vars(\{H,G\})$.

%$\mathbb D \models \forall \, (c_1 \vee \ldots \vee c_k \leftrightarrow d_1 \vee \ldots \vee d_m)$.

Now we show that $\Sigma$ is an $\mathcal A$-definable model of 
$P_{i+1}$ that is tight on $\textit{Defs}_{i+1}$.
By the hypothesis that  $\Sigma$ is an $\mathcal A$-definable model of 
$P_{i}$, the fact that $Y_1,\ldots,Y_r$ do not occur in $(G,H)$,  
and the distributivity law, we have that

$\mathbb D\models\forall(\exists Y_1\ldots\exists Y_r\ (c_1 \vee \ldots \vee c_k) \wedge \Sigma(G) \rightarrow \Sigma(H))$

%$\mathbb D\models\forall((c_1 \vee \ldots \vee c_k) \wedge \Sigma(G) \rightarrow \Sigma(H))$

\noindent
and hence 

$\mathbb D\models\forall(\exists Z_1\ldots\exists Z_s\ (d_1 \vee \ldots \vee d_m) \wedge \Sigma(G) \rightarrow \Sigma(H))$.

%$\mathbb D\models\forall((d_1 \vee \ldots \vee d_m) \wedge \Sigma(G) \rightarrow \Sigma(H))$.

\noindent
Thus, by using again the distributivity law, and the fact that
$Z_1,\ldots, Z_s$ do not occur in $(G,H)$,  
 we get that $\Sigma$ is an $\mathcal A$-definable model of $P_{i+1}$. Moreover, 
$\Sigma$ is tight on $\textit{Defs}_{i+1}$, which is equal to $\textit{Defs}_i$.\hfill
\end{proof}

\subsection*{Proof of Theorem \ref{thm:C-compl}}

\begin{proof}
Let us assume that there exists an $\mathcal A$-definable model $\Sigma'$ of $P_{i+1}$,
for $i\!=\!0,\ldots,n-1$. 
We will construct an \mbox{$\mathcal A$-definable model} $\Sigma$ of~$P_{i}$.
The proof proceeds by cases on the transformation rule applied to derive
$P_{i+1}$ from $P_i$.

\medskip
\noindent
({\it Case} R1) Suppose that $P_{i+1}$ is derived from $P_i$ by applying the
definition rule. Thus,  $P_{i+1}=P_i\cup \{D\}$, where $D$ is a new clause.
Let $\Sigma$ be equal to $\Sigma'$. $\Sigma$ is an $\mathcal A$-definable model of every subset of 
$P_{i+1}$, and hence it is an $\mathcal A$-definable model of~$P_i$.

\medskip
\noindent
({\it Case} R2) Suppose that $P_{i+1}$ is derived from $P_i$ by applying the
unfolding rule. Thus, $P_{i+1}=(P_i \setminus \{C\}) \cup 
\{H\leftarrow  c, {c}_j,G_1, B_j, G_2 \mid  j\!=\!1, \ldots, m\}$, where $C$ is
the clause $H\leftarrow c,G_1,p(X_1,\ldots,X_k),G_2$ in $P_i$
and $\{p(X_1,\ldots,X_k)\leftarrow {c}_j, B_j \mid  j\!=\!1, \ldots, m\}$ 
is the set of clauses in~$P_i$ whose head predicate is $p$.
%and, for $j\!=\!1, \ldots, m,$ $c\wedge c_j$ is $\mathbb D$-satisfiable.
From the hypothesis that the transformation is not a self-unfolding, 
that is,  $H$ is either \textit{false} or its predicate is different from $p$, it follows that the set
$\{p(X_1,\ldots,X_k)\leftarrow {c}_j, B_j \mid  j\!=\!1, \ldots, m\}$
is a subset of $P_{i+1}$. Since $\Sigma'$ is an $\mathcal A$-definable model of $P_{i+1}$, we have that,
for $j\!=\!1, \ldots, m,$

$\mathbb D \models  \forall(c \wedge {c}_j \wedge \Sigma'(G_1)\wedge \Sigma'(B_j)\wedge \Sigma'(G_2) \rightarrow \Sigma'(H))$   ~~and\hfill (1)\hspace{5mm}

$\mathbb D \models  \forall( {c}_j \wedge \Sigma'(B_j) \rightarrow \Sigma'(p(X_1,\ldots,X_k)))$ \hfill (2)\hspace{5mm}

\noindent
Now let us define

$\Sigma(q(X_1,\ldots,X_l)) =  \Sigma'(q(X_1,\ldots,X_l))$ \hspace{18mm} for $q$ different from $p$,
\hfill (3)\hspace{5mm}

$\Sigma(p(X_1,\ldots,X_k)) = \exists Y_1 \ldots\exists Y_n \bigvee_{j\!=\!1}^m 
({c}_j\wedge \Sigma'(B_j))$ \hfill (4)\hspace{5mm}

\noindent
where $\{Y_1, \ldots, Y_n\} = \textit{Fvars}(\bigvee_{j\!=\!1}^m 
({c}_j\wedge \Sigma'(B_j))) \setminus \{X_1,\ldots,X_k\}$.

\noindent
From (2) and (4) it follows that

$\mathbb D \models  \forall(\Sigma(p(X_1,\ldots,X_k)) \rightarrow \Sigma'(p(X_1,\ldots,X_k)))$ \hfill (5)\hspace{5mm}

\noindent
and hence, for every conjunction of atoms $G$,

$\mathbb D \models  \forall(\Sigma(G) \rightarrow \Sigma'(G))$ \hfill (6)\hspace{5mm}

\noindent
Now we show that $\Sigma$ is an $\mathcal A$-definable model of $P_i$, that is, $\Sigma$ is an $\mathcal A$-definable model of each clause~$D$ in $P_i$.
We consider the following three subcases.

\medskip
\noindent
({\it Subcase} 1) $D$ is the clause $C$: $H\leftarrow c,G_1,p(X_1,\ldots,X_k),G_2$ to which the unfolding rule is applied.
By definition of $\Sigma$, since $H$ is either \textit{false} or its predicate is different from $p$, 
by~(3), we get

$\Sigma(H)=\Sigma'(H)$.

\noindent
and hence, by (1), we get

$\mathbb D \models  \forall(c \wedge \Sigma'(G_1)\wedge \bigvee_{j\!=\!1}^m( {c}_j\wedge \Sigma'(B_j))\wedge \Sigma'(G_2) \rightarrow \Sigma(H))$

\noindent
By (4), we get

$\mathbb D \models  \forall(c \wedge \Sigma'(G_1)\wedge \Sigma(p(X_1,\ldots,X_k))  \wedge  \Sigma'(G_2) \rightarrow \Sigma(H))$

\noindent
and, finally, by (6),

$\mathbb D \models  \forall( 
c \wedge \Sigma(G_1)\wedge \Sigma(p(X_1,\ldots,X_k))  \wedge \Sigma(G_2) \rightarrow \Sigma(H))$

\medskip
\noindent
({\it Subcase} 2) $D$ is one of the clauses  $p(X_1,\ldots,X_k)\leftarrow {c}_j, B_j $ used for unfolding
$p(X_1,\ldots,X_k)$ in $C$. From the definition of $\Sigma(p(X_1,\ldots,X_k))$ given by (4), 
it follows that

$\mathbb D \models  \forall( {c}_j \wedge \Sigma'(B_j) \rightarrow \Sigma(p(X_1,\ldots,X_k)))$ 

\noindent
and hence, by (6),

$\mathbb D \models  \forall( {c}_j \wedge \Sigma(B_j) \rightarrow \Sigma(p(X_1,\ldots,X_k)))$ 

\medskip
\noindent
({\it Subcase} 3) $D$ is a clause in  $P_i \setminus (\{C\} \cup \{ p(X_1,\ldots,X_k)\leftarrow {c}_j, B_j \mid  j\!=\!1, \ldots, m\})$. Let~$D$ be a clause of the form $K \leftarrow e, Q$. Since $K$ is either \textit{false} or its 
predicate is different from~$p$, by~(3) it follows that $\Sigma(K) =  \Sigma'(K)$.
Since $D$ is different from $C$, it also belongs to $P_{i+1}$, and by the
hypothesis that $\Sigma'$ is an $\mathcal A$-definable model of $P_{i+1}$, we have that

$\mathbb D \models  \forall( e \wedge \Sigma'(Q) \rightarrow \Sigma(K))$ 

\noindent
Thus, by~(6),

$\mathbb D \models  \forall( e \wedge \Sigma(Q) \rightarrow \Sigma(K))$. 

\medskip
\noindent
({\it Case} R3) Suppose that $P_{i+1}$ is derived from $P_i$ by applying the
folding rule. Thus, \( P_{i+1}=(P_{i}\setminus\{C\})\cup \{E \} \), where
$C$ is the clause $H\leftarrow c, G_1,Q,G_2$ in $P_i$ and
$E$ is the clause \( H\leftarrow e, G_1, K\vartheta, G_{2} \) derived
%$E$ is the clause \( H\leftarrow c, G_1, K\vartheta, G_{2} \) derived
using  the clause $D$: $K \leftarrow d, B$ in $\textit{Defs}_i$
according to rule R3. 
Moreover, Conditions (i)--(iii) listed above when introducing rule R3, do hold.

Now we show that $\Sigma'$ is an $\mathcal A$-definable model of 
$P_{i}$, and hence we can take $\Sigma$ to be equal to $\Sigma'$.
From the hypothesis that the application of folding is reversible it follows
that $D$ belongs to $P_{i+1}$, and since $\Sigma'$ is an  $\mathcal A$-definable model of 
$P_{i+1}$, we have that

\( \mathbb D \models \forall (d \wedge \Sigma'(B) \rightarrow \Sigma'(K))\)  ~ and

\( \mathbb D \models \forall (e \wedge \Sigma'(G_1), \Sigma'(K\vartheta), \Sigma'(G_{2}) \rightarrow \Sigma'(H))\)

%\( \mathbb D \models \forall (c \wedge \Sigma'(G_1), \Sigma'(K\vartheta), \Sigma'(G_{2}) \rightarrow \Sigma'(H))\)

\noindent
and hence, by the definition of symbolic interpretation,

\( \mathbb D \models \forall (e \wedge d\vartheta \wedge \Sigma'(G_1), \Sigma'(B\vartheta), \Sigma'(G_{2}) \rightarrow \Sigma'(H))\)
%\( \mathbb D \models \forall (c \wedge d\vartheta \wedge \Sigma'(G_1), \Sigma'(B\vartheta), \Sigma'(G_{2}) \rightarrow \Sigma'(H))\)

\noindent
By Conditions (i)--(ii) of rule~R3, we get

\( \mathbb D \models \forall (c \wedge \Sigma'(G_1), \Sigma'(Q), \Sigma'(G_{2}) \rightarrow \Sigma'(H))\)

\noindent
and thus $\Sigma'$ is an $\mathcal A$-definable model of 
$P_{i}$.

\medskip
\noindent
({\it Case} R4) Suppose that $P_{i+1}$ is derived from $P_i$ by applying the
constraint replacement rule. Thus, $P_{i+1} = (P_i \setminus\{(H\leftarrow c_1, G), \ldots,(H\leftarrow c_k, G)\}) \cup \{(H\leftarrow d_1, G), $ $\ldots,(H\leftarrow d_m, G)\}$, where

$\mathbb D \models \forall \, (\exists Y_1\ldots\exists Y_r\ (c_1 \vee \ldots \vee c_k) \leftrightarrow \exists Z_1\ldots\exists Z_s\ (d_1 \vee \ldots \vee d_m))$,

\noindent
$\{Y_1,\ldots,Y_r\}=Fvars(c_1 \vee \ldots \vee c_k)\setminus vars(\{H,G\})$,
and $\{Z_1,\ldots,Z_s\}=Fvars(d_1 \vee \ldots \vee d_m)\setminus vars(\{H,G\})$.

Now we show that $\Sigma'$ is an $\mathcal A$-definable model of 
$P_{i}$, and hence we can take $\Sigma$ to be equal to $\Sigma'$.
By the hypothesis that  $\Sigma'$ is an $\mathcal A$-definable model of 
$P_{i+1}$, the fact that
$Z_1,\ldots, Z_s$ do not occur in $(G,H)$,  and the distributivity law, 
we have that

$\mathbb D\models\forall(\exists Z_1\ldots\exists Z_s\ (d_1 \vee \ldots \vee d_m) \wedge \Sigma'(G) \rightarrow \Sigma'(H))$

\noindent
and hence 

$\mathbb D\models\forall(\exists Y_1\ldots\exists Y_r\ (c_1 \vee \ldots \vee c_k) \wedge \Sigma'(G) \rightarrow \Sigma'(H))$.

\noindent
Thus, by using again the distributivity law and the fact that
$Y_1,\ldots, Y_r$ do not occur in $(G,H)$,  
we get that $\Sigma'$ is an $\mathcal A$-definable model of $P_{i}$.\hfill
\end{proof}

%\medskip
%\noindent
%({\it Case} R4) Suppose that $P_{i+1}$ is derived from $P_i$ by applying the
%constraint replacement rule. Thus, $P_{i+1}\! =\! (P_i\! -\!\{(H\!\leftarrow \!
%c_1,G), \ldots,(H\!\leftarrow\! c_k,G)\}) \cup \{(H \!\leftarrow \! d_1,G), $  $\ldots,(H \!\leftarrow \! d_m,G)\}$, where
%$\mathbb D \models \forall \, (c_1 \vee \ldots \vee c_k \leftrightarrow d_1 \vee \ldots \vee d_m)$.
%
%Now we show that $\Sigma'$ is an $\mathcal A$-definable model of 
%$P_{i}$, and hence we can take $\Sigma$ to be equal to $\Sigma'$.
%By the hypothesis that  $\Sigma$ is an $\mathcal A$-definable model of 
%$P_{i+1}$ and the distributivity law, we have that
%
%$\mathbb D\models\forall((d_1 \vee \ldots \vee d_m) \wedge \Sigma'(G) \rightarrow \Sigma'(H))$
%
%\noindent
%and, by the equivalence $\mathbb D \models \forall \, (c_1 \vee \ldots \vee c_k \leftrightarrow d_1 \vee \ldots \vee d_m)$, we get
%
%
%$\mathbb D\models\forall((c_1 \vee \ldots \vee c_k) \wedge \Sigma(G) \rightarrow \Sigma(H))$ 
%
%\noindent
%Thus, by using again the distributivity law, we get that
%$\Sigma'$ is an $\mathcal A$-definable model of $P_{i}$.

\end{document}